\documentclass[a4paper,11pt]{article}

\usepackage{jheppub} 
\usepackage[bottom]{footmisc}
\usepackage{amssymb}
\usepackage{amsmath}
\usepackage{amsthm}
\usepackage[usenames,dvipsnames]{xcolor}
\usepackage{epsfig}
\usepackage{dcolumn}
\usepackage{tikz}
\usetikzlibrary{shapes.geometric, arrows}
\usepackage{upgreek}
\usepackage{setspace}
\usepackage{enumerate}
\usepackage[inline]{enumitem}

\usepackage{array,multirow,bigdelim,arydshln}
\usepackage{appendix}
\usepackage{xparse}
\usepackage{stmaryrd}
\usepackage[T1]{fontenc} 
\usepackage{mathtools}
\usepackage{physics} 
\usepackage{adjustbox}
\usepackage{multirow}
\usepackage{graphicx} 
\usepackage{float} 
\graphicspath{{./images/}}
\usepackage[nottoc]{tocbibind}
\usepackage{hyperref}
\usepackage[utf8]{inputenc}
\hypersetup{
	colorlinks,
	urlcolor=Maroon,
	linkcolor=Maroon,
	citecolor=Maroon
	}

\NewDocumentCommand{\binomial}{omm}
 {%
  \genfrac(){0pt}{}{#2}{#3}%
  \IfValueT{#1}{_{\!#1}}%
 }
\NewDocumentCommand{\eulerian}{omm}
 {%
  \genfrac<>{0pt}{}{#2}{#3}%
  \IfValueT{#1}{_{\!#1}}%
 }

\def \s {\sigma}

\usepackage{latexsym}
\usepackage{tikz}

\usetikzlibrary{shapes.geometric, arrows,decorations}
\usetikzlibrary{decorations.markings}
\usetikzlibrary{patterns}

\theoremstyle{plain}

\theoremstyle{definition}

\def\bea#1\ea{\begin{eqnarray}#1\end{eqnarray}}
\def\be#1\ee{\begin{equation}#1\end{equation}}
\def\ba#1\ea{\begin{align}#1\end{align}}

\def\nl{\nonumber\\}
\def\non{\nonumber}

\def\yz#1\yz {{\color{blue} [[YZ: #1]] }}
\def\yqz#1\yqz {{\color{red} [[YQZ: #1]] }}

\usepackage{amsmath}
\usepackage{multicol}
\usepackage{bbm}
\usepackage{enumerate}

\usepackage{amsthm}
\usepackage{mathrsfs}
\usepackage{upgreek}
\usepackage{amssymb}
\usepackage{bm}
\usepackage{setspace}
\usepackage{array,multirow,arydshln}
\usepackage{bigdelim}
\usepackage{scalerel}
\usepackage{diagbox}
\usepackage{caption}
\usepackage{tabularx}

\usepackage{tikz}

\usetikzlibrary{shapes.geometric,arrows,arrows.meta,decorations.pathmorphing,decorations.markings,patterns}

\def\<{\langle}
\def\>{\rangle}
\def\a{\alpha}
\def\b{\beta}

\def\e{\epsilon}

\def\s{\sigma}

\def\beq{\begin{equation}}
\def\eeq{\end{equation}}

\usepackage{enumitem}

\usepackage{enumerate}

\usepackage[percent]{overpic}
\usepackage{multirow}
\usepackage{slashed}

\DeclareMathOperator{\sgn}{sgn}

\usepackage{subfig}
\usepackage{float}

\def\beq{\begin{equation}}
\def\eeq{\end{equation}}

\usepackage{enumitem}

\usepackage{enumerate}

\DeclareFontFamily{U}{mathx}{\hyphenchar\font45}
\DeclareFontShape{U}{mathx}{m}{n}{
      <5> <6> <7> <8> <9> <10>
      <10.95> <12> <14.4> <17.28> <20.74> <24.88>
      mathx10
      }{}
\DeclareSymbolFont{mathx}{U}{mathx}{m}{n}
\DeclareFontSubstitution{U}{mathx}{m}{n}
\DeclareMathAccent{\widecheck}{0}{mathx}{"71}

\title{One-loop diagrams with quadratic propagators from the worldsheet}

\date{\today}

\author[a,b,c,d]{Bo Feng,}

\author[e,f,d]{Song He,}

\author[g]{Yong Zhang}

\author[e,h]{and Yao{-}Qi Zhang}%

\affiliation[a]{Beijing Computational Science Research Center, Beijing 100084, China}
\affiliation[b]{Zhejiang Institute of Modern Physics, Zhejiang University, Hangzhou, 310027, P. R. China }
\affiliation[c]{Center of Mathematical Science, Zhejiang University, Hangzhou, 310027, P. R. China}
\affiliation[d]{Peng Huanwu Center for Fundamental Theory, Hefei, Anhui 230026, China}
\affiliation[e]{CAS Key Laboratory of Theoretical Physics, Institute of Theoretical Physics, Chinese Academy of Sciences, Beijing 100190, China}
\affiliation[f]{
School of Fundamental Physics and Mathematical Sciences, Hangzhou Institute for Advanced Study, UCAS;
International Centre for Theoretical Physics Asia{-}Pacific, Beijing/Hangzhou, China}

\affiliation[g]{Perimeter Institute for Theoretical Physics, Waterloo, ON N2L 2Y5, Canada}

\affiliation[h]{School of Physical Sciences, University of Chinese Academy of Sciences, No.19A Yuquan Road, Beijing 100049, China}

\emailAdd{fengbo@csrc.ac.cn}

\emailAdd{songhe@itp.ac.cn}

\emailAdd{yzhang@perimeterinstitute.ca}

\emailAdd{zhangyaoqi@itp.ac.cn}

\date{\today}

\abstract{It is well known that forward limits of tree-level amplitudes (and those trivalent diagrams they consist of) produce one-loop amplitudes and trivalent diagrams with propagators linear in the loop momentum. They naturally arise from one-loop worldsheet formulae, and an important open problem is how to recombine them into usual one-loop diagrams with quadratic propagators. In this paper, we study a new collection of worldsheet functions: generalized one-loop Parke-Taylor factors with tensor numerators, which are conjectured to serve as a basis for one-loop worldsheet functions with this nice property. We present all-multiplicity, closed-form expressions for combinations of one-loop trivalent diagrams with quadratic propagators and tensor numerators to arbitrary rank (including possible tadpole contributions), produced by any pair of Parke-Taylor factors.  We also briefly comment on reducing worldsheet functions onto such a basis, and applications to one-loop amplitudes in physical theories.
}

\begin{document}

\maketitle


\addtocontents{toc}{\protect\setcounter{tocdepth}{2}}


\section{Introduction}
Recent years have witnessed enormous progress in understanding novel structures and symmetries of scattering amplitudes in various theories, as well as various methods for calculating amplitudes beyond the traditional Feynman-diagram approach. One such approach is the so-called Cachazo-He-Yuan (CHY) formalism~\cite{Cachazo:2013gna,Cachazo:2013hca,Cachazo:2013iea}, inspired by representations of tree-level amplitudes as integrals over moduli space of certain maps from $n$-punctured sphere $\mathcal{M}_{0,n}$
~\cite{Witten:2003nn,Roiban:2004vt,Cachazo:2012da,Cachazo:2012kg,Huang:2012vt,Cachazo:2013iaa}. For $n$-point massless particles in general $D$ dimensions, the key point of CHY formalism lies at the scattering equations~\cite{Cachazo:2013gna}
\begin{equation}
E_a:=\sum_{b \neq a} \frac{2 \,k_{a} \cdot k_{b}}{\sigma_{a}-\sigma_{b}}=0, \quad a \in\{1,2, \ldots, n\},
\end{equation}
where $k_a$ is the momentum carried by particle $a$ and $\sigma_a$ denotes the position of the $a$-th puncture. The tree-level scattering amplitude admits CHY representation
\begin{equation}
\label{treechy}
M _{n}^{\text {tree }}=\int \frac{d^{n} \sigma}{\operatorname{vol SL}(2, \mathbb{C} )} \prod_{a} {}^{^{\prime}} \delta\left(E_a\right) I _{n}(\{k, \epsilon, \sigma\}),
\end{equation}
where the CHY integrand $I_n$ depends on the physical theories~\cite{Cachazo:2013hca,Cachazo:2013iea,Cachazo:2014nsa,Cachazo:2014xea}. 
Not only do CHY formulae provide a new, universal representation of scattering amplitudes in various theories, but they also offer insights into new relations between them; among others, Kawai-Lewellen-Tye (KLT) relations~\cite{Kawai:1985xq,Bjerrum-Bohr:2010pnr} and  Bern-Carrasco-Johansson (BCJ) double copy (based on color-kinematics duality) for gauge theories and gravity~\cite{loopBCJ,Bern:2010fy,Adamo:2022dcm} become manifest in CHY~\cite{Cachazo:2013gna,Cachazo:2013iea}, which have been extended to relations among various effective field theories ~\cite{Cachazo:2014xea,Cheung:2014dqa}.  


It has been understood that what underpins CHY formulae is ambitwistor string theory~\cite{Mason:2013sva,Geyer:2022cey}, which gives elegant worldsheet models with worldsheet correlators and beautifully produces CHY (half-)integrands such as Parke-Taylor factor and Pfaffian. These worldsheet models can be naturally extended to loop level:
we expect that $g$-loop amplitudes can be expressed as integrals over moduli space of genus $g$, at least for $g\leq 2$
~\cite{Adamo:2013tsa,Geyer:2015bja,Geyer:2015jch,Geyer:2016wjx,Geyer:2018xwu,Geyer:2019hnn,Geyer:2021oox}. More specifically, Adamo, Casali, and Skinner proposed a conjecture for the one-loop formula of supergravity amplitudes based on scattering equations on a torus~\cite{Adamo:2013tsa}, and Geyer {\it et al}  transformed it into integrals over punctured nodal Riemann spheres~\cite{Geyer:2015bja}.  The infinite-tension limit of conventional strings is also found to be able to give CHY integrands in a direct way~\cite{Berkovits:2013xba,Adamo:2015hoa}, whose loop momenta at higher genus can be introduced by chiral splitting \cite{DHoker:1988pdl, DHoker:1989cxq}. Alternatively, one loop CHY formulae can also be obtained from the forward limit of tree-level ones with two massive particles~\cite{He:2015yua,Cachazo:2015aol} akin to the Q-cuts of \cite{Baadsgaard:2015twa}. One-loop CHY becomes useful for deriving KLT-like relations {\it etc.} at one-loop level~\cite{He:2016mzd,He:2017spx}.



Although ambitwistor-string/CHY formulae at one-loop level (for gauge theories and gravity {\it etc.})
have been proposed for a while, notably they usually only produce one-loop integrands and trivalent diagrams with propagators linear in loop momenta (except for an overall $1/\ell^2$), which are difficult to be integrated (see {\it c.f.} \cite{Geyer:2015bja,Geyer:2015jch,Cachazo:2015aol,Geyer:2017ela,He:2017spx,Edison:2020uzf}).  Even though lower-point quadratic propagators could be obtained by brute-force manipulation of linear propagators, such as what have
done in \cite{Porkert:2022efy} for one-loop four-point amplitudes in (half-)maximally supersymmetric Einstein-Yang-Mills (EYM) theory, a general method for all multiplicities is still missing despite lots of efforts ~\cite{Cardona:2016bpi,Cardona:2016wcr,Gomez:2017lhy, Farrow:2020voh,Gomez:2016cqb,Gomez:2017cpe,Ahmadiniaz:2018nvr,Agerskov:2019ryp,Baadsgaard:2015hia}.
 In a recent paper~\cite{Edison:2021ebi}, it was shown that the field-theory limits of homology invariants \cite{Mafra:2017ioj,Mafra:2018pll, DHoker:2020prr}, which can be viewed as certain kinds of genus-one string correlators,  can act as one-loop CHY  integrands that produce quadratic propagators. The limit of homology invariants is relatively simple and more general half-integrands may be able to be expanded onto them \cite{Mafra:2018nla,Mafra:2018qqe}. So their discovery simplifies the study of quadratic propagators.
 In this note, we continue the research of one-loop CHY integrands that can give quadratic propagators from a different perspective (in the hope that these two directions will converge in the future).

Our starting point is
the one-loop Parke-Taylor
factor
\ba \label{onelooppt}
{\rm PT}(1,2,\cdots,n):=  \sum_{i=1}^n{\rm PT}^{\rm tree}(+,i,i+1,\cdots,n,1,2,\cdots, i-1,-)\,,
\ea
as the sum of $n$ tree PT factors, which was first proposed in \cite{Geyer:2015bja}. Denoting the loop momentum as $\ell$,    $+$, and $-$ in the above formula refer to the legs $\pm \ell$ in which the forward limit is taken as we will explain later. There is no doubt that the one-loop CHY integral of such scalar PT factors including the relabelling of \eqref{onelooppt} just produces one-loop trivalent diagrams with trivial kinematic numerators \cite{He:2015yua}. 
We find it very useful to introduce the following generalizations of the one-loop Parke-Taylor
factors
 \ba
\label{ptvectorg}
{\rm PT}^{\mu_1,\mu_2,\cdots,\mu_r}(1,2,\cdots,n):=  \sum_{i=1}^n {\rm PT}^{\rm tree}(+,i,i+1,\cdots, i-1,-)
\prod_{j=1}^r (\ell^{\mu_j}- k_{12\cdots i-1}^{\mu_j})\,;
\ea
 obviously, the $r$ Lorentz indices for  rank-$r$ generalized PT factors are totally symmetric. In the following, we will show how
these $\ell$-decorated PT factors produce loop integrands, which consist of trivalent diagrams with quadratic propagators and $\ell$-dependent numerators. 
From now on, a (generalized) PT factor means a one-loop generalized PT factor, as the main object of our paper.

In all cases we have studied, we find that string-inspired integrands proposed in \cite{Edison:2021ebi}, which produce diagrams with quadratic propagators, can always be expanded as linear combinations of our PT factors. Besides, although string-inspired integrands are known explicitly for certain multiplicities, finding all string-inspired integrands at higher points needed for one-loop amplitudes becomes more and more challenging \cite{Mafra:2018pll}. On the other hand, our definition of generalized PT factors holds for arbitrary multiplicities; once we find patterns of trivalent diagrams with quadratic propagators produced by them, they can be used for arbitrary multiplicities without any difficulty.

As we will show, up to rank-$r$ tensor, there are $(r+1)(n-1)!$ linearly independent one-loop generalized PT factors at algebraic level. On the support of one-loop scattering equations, there are further relations among them, as we will comment later. We have strong evidence that such generalized PT factors can serve as a basis of one-loop CHY half-integrands that produce quadratic propagators. It would be highly desirable to have proof of this statement. From now on, we would just use  {\it {quadratic half-integrands}} to refer to those that produce quadratic propagators \footnote{Here we mean their CHY integrals with themselves or any other confirmed quadratic half-integrands including all generalized PT factors produce loop integrands only involving quadratic propagators.  }: in this sense, we conjecture that the rank of quadratic half-integrands up to rank $r$ is no more than $(r+1)(n-1)!$, either. 
As concrete evidence for this conjecture, we show a very non-trivial kind of expansion in the latter part of this paper.

In general, our strategy  towards  one-loop diagrams with quadratic propagators is as follows: one first decomposes general quadratic half-integrands as linear combinations of PT factors, which reduces general one-loop CHY integrals into those with two PT factors, just as at the tree level;
in this way, we will have a representation of one-loop integrands involving quadratic propagators only. We will present an explicit application of our general idea to amplitudes in physical theories such as super-Yang-Mills (SYM) and supergravity (SG).

The rest of the paper is organized as follows. We first review one-loop CHY formulae in section \ref{sec:rev}, and introduce these generalized PT factors in a natural way in section \ref{sec:gen}. After revisiting the CHY integrals of scalar PT factors in section \ref{sec:inte1}, we propose the closed-form formula for generalized PT factors in section \ref{sec:inte2}, with two relevant proofs  present in appendix \ref{proofpro} and  \ref{tadpolesim}. 
Section \ref{subcycle} is dedicated to a non-trivial kind of expansion onto the PT basis. In section \ref{applicationsec},  we apply our results to amplitudes in SYM and SG with half-maximal supersymmetry. 
Finally, we conclude with a brief summary and outlook in section \ref{sec:con}.  Two more appendices \ref{twomostgen} and \ref{appeappde} complement the material in the main text.

\section{
\label{sec:rev}Review of one-loop CHY formulae}

One-loop CHY formulae express loop integrands with $n$ external in-going massless  legs as integrals over the moduli space of 
degenerated tori, i.e.,  
nodal Riemann spheres  localized by the universal one-loop  scattering equations (SE) \cite{Geyer:2015bja,Geyer:2015jch}, which can 
  be obtained from  $(n+2)$-pt massless tree-level CHY formulae \eqref{treechy} in the forward limit in higher dimensions 
  \cite{He:2015yua,Cachazo:2015aol},
\ba
\label{oneloopchyori}
\mathcal{M}(\ell)=\frac{1}{\ell^2}
 \lim_{K_-\to-K_+}
\int
\frac{1 }{{\rm SL}(2,\mathbb{C})}
\prod_{a=1}^{ n, +,-} 
 \mathrm{~d} \sigma_{a} 
\prod_{a=1}^{ n, +,-} 
  {}'\delta
  \!
\left(
\sum_{b=1  \atop b \neq a}^{ n, +,-} \frac{ 2\, K_a\cdot K_b }{\sigma_{a b}}
\right)
\!
{\hat {\mathcal{I}}}_{L}(\ell) {\hat {\mathcal{I}}}_{R}(\ell)\,,
\ea
where  $\sigma_{ab}:= \sigma_a-\sigma_b$ and the half-integrands ${\hat {\mathcal{I}}}_{L/R}(\ell) $ depend on theories.
\footnote{In principle,  the one-loop CHY integrands could be a linear combination of  ${\hat {\mathcal{I}}}_{L}(\ell) {\hat {\mathcal{I}}}_{R}(\ell)$. However, 
for many theories such as (supersymmetric) YM, GR, and EYM, their CHY integrands factorize. In this paper, we focus on this kind of CHY integrands. }
The one-loop amplitudes are then given by
\ba 
\label{loopintegral}
\pmb{\mathcal{M}}=\int \mathrm{d}^{D} \ell \, \mathcal{M}(\ell)\,.
\ea
In \eqref{oneloopchy}, the on-shell momenta $K_a$ in higher dimensions are related to those $k_i$ in original dimensions and off-shell loop momentum $\ell$ via
\ba
\label{higherkin}
K_{+}\to {\cal L}=(\ell, |\ell|),\qquad  K_{-}\to -{\cal L}=(-\ell, -|\ell|) , \qquad K_i=(k_i,0) \quad \forall 1\leq i\leq n\,,
\ea
which are subject to  $K_{+}+K_{-}+\sum_{i=1}^n K_i=0$. In the forward limit $K_-\to - K_+$,  we have {\it the momentum conservation of $n$ external legs} which we denote as MCEL,
\ba
\label{defmc}
{\rm MCEL}: \quad \sum_{i=1}^n k_i \to 0\,.
\ea

Just as at tree level \eqref{treechy},
 three punctures in \eqref{oneloopchyori} can be fixed by using 
the ${\rm SL}(2,{\mathbb C})$ gauge redundancy in the nodal Riemann spheres and three  SEs can be removed by complementing proper factors.  A common way is to fix the two nodal punctures as %
$\sigma_+\to 0, \sigma_-\to \infty$. One more puncture can be fixed  as well  using ${\rm SL}(2,{\mathbb C})$, say $\s_1\to 1$, but we keep it general nevertheless when we focus on the half-integrands. This way, the one-loop 
CHY formulae simplify to
\footnote{Even though \eqref{oneloopchy} is simpler,  the original formula \eqref{oneloopchyori} is more general. Throughout the paper, we may use the gauge fixing $\sigma_+\to 0, \sigma_-\to \infty$ to simplify some one-loop CHY formulae by keeping in mind that  their  ${\rm SL}(2,{\mathbb C})$ redundancy can be recovered directly and their CHY integrals are actually performed by using \eqref{oneloopchyori}. }
\ba
\label{oneloopchy}
\mathcal{M}(\ell)=\frac{1}{\ell^2}
\int
\underbrace{\prod_{i=2}^{n} \mathrm{~d} \sigma_{i} \, \delta
\overbrace{
\left(
\frac{2\,\ell\! \cdot\! k_{i} }{\sigma_{i}}+\sum_{j=1 \atop j \neq i}^{n} \frac{s_{ij} }{\sigma_{i j}}
\right)
}
^
{\rm SE}
}_{:=d\mu_n}
\!
{\mathcal{I}}_{L}(\ell) {\mathcal{I}}_{R}(\ell)\,.
\ea
Throughout this work, our conventions for Mandelstam invariants $s_{12 \ldots p}$ and multiparticle momenta $k_{12 \ldots p}$ in original dimensions are as follows:
\ba
k_{12 \ldots p} := k_{1}+k_{2}+\ldots+k_{p}, \quad s_{12 \ldots p} := 2\sum_{i<j}^{p} k_{i} \cdot k_{j}, \quad s_{12 \ldots p, \pm \ell} := 2\sum_{i<j}^{p} k_{i} \cdot k_{j} \pm 2 \ell \cdot k_{12 \ldots p}.
\ea
 Besides, we use the shorthand
 \ba
 \ell_{12\cdots p}:=  \ell+k_{12\cdots p}
 \ea
 such that  $\ell_{12\cdots p}^2$  denotes the quadratic propagator. As comparison, 
  $s_{12\cdots p,\ell}= \ell_{12\cdots p}^2-\ell^2  $ denotes linear propagator. Note that ${\cal L}^2=0,  ({\cal L}+K_I)^2= 2 {\cal L}\cdot K_I+ K_I^2= s_{I,\ell}$.  Hence
 the poles  $s_{I,\ell}$ including $s_{i,\ell}= 2\, \ell \!\cdot\! k_i$ linear in loop propagators can be expressed as poles of massless particles in higher dimensions. 
 
 For physical theories, of course, the $n$ external legs satisfy strict momentum conservation. However, we would like to treat \eqref{defmc} as a limit, which allows us to regularize $``0/0"$ indeterminate terms. Hence,  we may meet 
 naively 
divergent propagators like
 $s_{12\cdots,n-1}^{-1}$ in a massless bubble graph or even  $s_{12\cdots n}^{-1}$ in a tadpole graph in our formulae of this paper,
 %
 %
 which themselves won't directly produce loop integrands of physical theories but the combinations of which could. In the latter case,  the particular combinations should lead to some vanishing terms in the numerators as well that cancel the 
divergent propagators mentioned before.  A typical procedure used in  half- or quarter- maximal supersymmetric  theories is  Minahaning  \cite{Minahan:1987ha,Berg:2016wux,Berg:2016fui} where the momentum conservation of external legs is relaxed as $\sum_{i=1}^n k_i=p$ with $ p\cdot v=0 \, \,\forall v\neq k_1,\cdots,k_n$
 \footnote{This implies $p^2=0,p\cdot \ell=0$ and 
 $p\cdot \e_i=0$ with $\e_i$ denoting a polarization vector.} first. Once all divergent propagators in the external bubbles are canceled, one can set $p=0$ again.
 \footnote{Another way to regularize the $``0/0"$ indeterminate terms is just to throw away all the singular solutions of the one-loop SEs in \eqref{oneloopchyori},  which prevents the appearance of divergent propagators at the very beginning and should work for most general cases, including those without supersymmetry \cite{Cachazo:2015aol}.  However, such operations will lead to some non-trivial deformations of our main results in this paper, which we leave to future work.} Our paper focuses on $\ell$-dependent quadratic propagators and we will leave the divergent sub-tree propagators   untouched. \footnote{As shown in appendix \ref{proofpro}, our main results like \eqref{twogeneral} rely on the exact momentum conservation of $n+2$ points. MCEL is only necessary when we transfer linear loop propagators as quadratic ones. But we have never used MCEL  to simplify the sub-tree propagators, i.e., one can apply the MCEL later in his own way to deal with all divergent sub-tree propagators. }
  
  Since the  one-loop CHY integrals  \eqref{oneloopchyori} are actually tree-level ones, we can evaluate them by
 abundant methods at tree level \cite{Cachazo:2013iea,Cachazo:2015nwa,Baadsgaard:2015voa,Baadsgaard:2015ifa,Huang:2015yka,Bosma:2016ttj,Cardona:2016gon,Lam:2016tlk,Cachazo:2019aby,Edison:2020ehu,Gao:2017dek,He:2021lro}. 
However, in this way we just get plenty of linear propagators $s_{I,\ell}$, which makes the loop integral \eqref{loopintegral} difficult.
 To obtain quadratic propagators just as those in Feynman diagrams, we will review how they are related to linear propagators first.

\subsection{Polygons with linear and quadratic propagators}
As schematically shown below, one can use partial fraction and shifts of loop momenta to write the integral of scalar $n$-gon as the sum of $n$ terms \cite{Geyer:2015jch, Geyer:2015bja}: 
\ba \label{cutcut}
\raisebox{-.7cm}{
\begin{tikzpicture} [scale=0.75, line width=0.30mm]
\begin{scope}[xshift=-0.8cm]
\draw (0.5,0)--(-0.5,0);
\draw (-0.5,0)--(-0.85,-0.35);
\draw [dashed](-0.85,-0.35)--(-1.2,-0.7);
\draw (0.5,0)--(1.2,-0.7);
\draw[dashed] (-1.2,-1.7)--(-1.2,-0.7);
\draw (1.2,-1.7)--(1.2,-0.7);
\draw (1.2,-1.7)--(0.85,-2.05);
\draw[dashed] (0.85,-2.05)--(0.5,-2.4);
\draw[dashed] (-0.5,-2.4)--(0.5,-2.4);
\draw[dashed] (-0.5,-2.4)--(-1.2,-1.7);
\draw (-0.5,0)--(-0.7,0.4)node[left]{$n$};
\draw (0.5,0)--(0.7,0.4)node[right]{$1$};
\draw (1.2,-0.7)--(1.6,-0.5)node[right]{$2$};
\draw (1.2,-1.7)--(1.6,-1.9)node[right]{$3$};
\draw (0,0) node{$| \! |$};
\draw (-0.25,0.2)node{$-$};
\draw (0.25,-0.2)node{$+$};
\end{scope}
\draw[-> ](2.2,-1.2)  -- (3.7,-1.2);
\begin{scope}[xshift=2.7cm, yshift=0.5cm]
\draw(11.4,-2)node{$+ \ {\rm cyclic}(1,2,\ldots,n)$};
\draw (2.9,-2)node[left]{$+\ell$} -- (5.8,-2);
\draw (7.2,-2) -- (8.1,-2)node[right]{$-\ell$};
\draw (3.5,-2) -- (3.5,-1.5)node[above]{$1$};
\draw (4.5,-2) -- (4.5,-1.5)node[above]{$2$};
\draw (5.5,-2) -- (5.5,-1.5)node[above]{$3$};
\draw[dashed] (5.8,-2) -- (7.2,-2);
\draw (7.5,-2) -- (7.5,-1.5)node[above]{$n$};
%
\end{scope}
\end{tikzpicture}}\,,
\ea
where we have used partial fraction
\ba
\frac{1}{\prod_{i=1}^{m} D_{i}}=\sum_{i=1}^{m} \frac{1}{D_{i} \prod_{j \neq i}\left(D_{j}-D_{i}\right)}
\ea
to write a product of quadratic propagators in terms of one quadratic propagator and the remaining linearized ones; then using shifts of loop momenta, {\it e.g.}
\ba \label{shiftofloop}
\frac{1}{(\ell^2-\ell_{1}^2) \ell_{1}^2 (\ell_{12}^2-\ell_{1}^2) \cdots (\ell_{12\cdots ,n-1}^2-\ell_{1}^2)}
{\overset{\ell\to \ell-k_1}{\to}} &
\frac{1}{(\ell_{23\cdots n}^2-\ell^2) \ell^2 (\ell_{2}^2-\ell^2) \cdots (\ell_{23\cdots ,n-1}^2-\ell^2)}
\nl
=\,\,&
\frac{1}{\ell^2\, s_{2,\ell}s_{23,\ell}\cdots s_{23\cdots n,\ell}}\,,
\ea
the only quadratic propagator can be written as an overall factor of $1/\ell^2$. Since the shift \eqref{shiftofloop} does not affect loop integral \footnote{The problem of anomaly can be considered separately, which is beyond the scope of this paper.}, we have an equivalence relation:
\ba
\label{oneloopptint}
\frac{1}{\ell^2 \ell_{1}^2 \ell_{12}^2 \cdots \ell_{12\cdots ,n-1}^2}
\cong
& \frac{1}{\ell^2\, s_{1,\ell}s_{12,\ell}\cdots s_{12\cdots ,n-1,\ell}}+{\rm cyc}(1,2,\cdots,n)
\,;
\ea
throughout the paper, we will use $\cong$ to indicate that two integrands give the same result after integration (but they are not equal to each other at the integrand level).  Each term on the RHS of the above equation can be thought of as an $(n+2)$-pt massless tree-level trivalent graph with kinematics $ K_a$ explained before. This picture shows how an $n$-gon is related to $n$ trees in the ``forward limit''~\cite{He:2015yua,Cachazo:2015aol}.

Recall that the first example of a one-loop CHY formula that gives quadratic propagators is the $n$-gon formula with a scalar
PT factor \eqref{onelooppt} and the most symmetric factor
\begin{equation}\label{symmetricfactor}
\frac{(-1)^{n+1}}{\sigma_{1} \sigma_{2} \cdots \sigma_{n}}=\sum_{\pi \in S_{n-1}} {\rm PT} (1, \pi(2,3, \cdots, n))=\sum_{\rho \in S_{n}} {\rm PT} ^{\text {tree }}(+, \rho(1,2, \cdots, n),-)\,,
\end{equation}
where the tree-level PT factor of $(n{+}2)$ points is defined as:
\begin{equation}
\label{treept}
{\rm PT} ^{\text {tree }}(+, 1,2, \cdots, n,-):=\frac{1}{\sigma_{1} \sigma_{12} \sigma_{23} \cdots \sigma_{n-1, n}}\,.
\end{equation}

Each tree-level PT factor on the RHS of \eqref{onelooppt} together with the most symmetric factor produces a trivalent graph on the RHS of \eqref{oneloopptint}. Hence a cyclic sum of tree-level PT factors, {\it i.e.} the one-loop PT factor, produces an $n$-gon~\cite{Geyer:2015bja,Geyer:2015jch,He:2015yua,Feng:2016msc},
\begin{equation}\label{ngonappear}
\frac{1}{\ell^{2}} \int d \mu_{n} {\rm PT} (1,2, \cdots, n) \frac{(-1)^{n+1}}{\sigma_{1} \sigma_{2} \cdots \sigma_{n}} \cong \operatorname{gon}(1,2, \cdots, n)=\frac{1}{\ell^{2} \ell_{1}^{2} \ell_{12}^{2} \cdots \ell_{12 \cdots, n-1}^{2}} .
\end{equation}

Throughout the paper, it will be convenient to denote the  $n$-gon in \eqref{oneloopptint} as ${\rm gon}(1,2,\cdots,n)$ and in general we define
\ba
\label{partialgeneral}
{\rm gon}(A_1,A_2,\cdots,A_m):=\,\,& \frac{1}{\ell^2 \ell_{A_1}^2 \ell_{A_{12}}^2 \cdots \ell_{A_{12\cdots ,m-1} }^2}
\\
\cong \,\,&
 \frac{1}{\ell^2\, s_{A_1,\ell}s_{A_{12},\ell}\cdots s_{A_{12\cdots ,m-1},\ell}}+{\rm cyc}(A_1,A_2,\cdots,A_m)
\,,
\non
\ea
where $\ell_{A_{12}}= \ell_{A_{1}, A_{2}}= \ell+k_{A_1}+k_{A_2},\, s_{A_{12},\ell}= s_{A_{1}, A_{2},\ell}$ and    $A_1\sqcup A_2\sqcup \cdots \sqcup A_m=\{1,2,\cdots,n\}$.

We comment that whenever we talk about shifts of loop momentum, it is assumed that MCEL can be used to
change the representation of loop propagators, for example, $(\ell-k_{1})^2 \to (\ell+k_{{23\cdots n}})^2 $ in \eqref{shiftofloop}.

Next, let us quickly mention the relations of such polygons by shifts. By construction, we have the symmetry
\ba
{\rm gon}(A_1,A_2,\cdots,A_m)\cong {\rm gon}(A_2,\cdots,A_m,A_1)\,,
\ea
which are related by the shift of loop momentum $\ell\to \ell-k_{A_1}$.


When there are $\ell$-dependent numerators in the one-loop trivalent graphs, in addition to \eqref{partialgeneral}, we have the following equivalent relations based on partial fraction and  shifts of loop momenta,
\ba
&\label{partialgeneral2}
N(\ell)\, {\rm gon}(A_1,A_2,\cdots,A_m)
\\
\non
\cong\,\,&
 \frac{N(\ell)}{\ell^2\, s_{A_1,\ell}s_{A_{12},\ell}\cdots s_{A_{12\cdots ,m-1},\ell}}
 +
  \frac{N(\ell-k_{A_1})}{\ell^2\, s_{A_2,\ell}s_{A_{23},\ell}\cdots s_{A_{2\cdots ,m},\ell}}
 + \cdots
   +
  \frac{N(\ell-k_{A_{12\cdots m-1} })}{\ell^2\, s_{A_m,\ell}s_{A_{m1},\ell}\cdots s_{A_{m1\cdots ,m-2},\ell}}
 \,.
\ea
The observation is crucial since the one-loop CHY formulae \eqref{oneloopchy} would produce the results on the RHS of \eqref{partialgeneral2} at first. 
Then a good organization of them combines into quadratic propagators.
Constructing an integrand that produces each term on the RHS of \eqref{partialgeneral2} is relatively simple based on the results of the tree level.  However, constructing an integrand that produces such a particular combination of all the terms on the RHS is challenging.  Our essential task 
is to construct well-organized minimal blocks, such that as long as a one-loop CHY integrand
can be written as the combination of these blocks,  it produces quadratic propagators automatically.

\section{Generalized PT factors
\label{sec:gen}}

To produce an integrand with $\ell$-dependent numerator, for example, $\ell ^\mu\, {\rm gon}(1,2,\cdots,n) $, one might naively choose $\ell^\mu\, {\rm PT}(1,2,\cdots,n)$, but this is wrong since it has not taken into account the effect of shifting loop momenta \eqref{shiftofloop}.
However, inspired by the fact,
\ba
\label{numeratorngon}
\frac{N(\ell^\mu)}{\ell^2 \ell_{1}^2  \cdots \ell_{12\cdots ,n-1}^2}
\cong &
 \frac{N(\ell^\mu)}{\ell^2\, s_{1,\ell}\cdots s_{12\cdots ,n-1,\ell}}
 + \frac{N(\ell^\mu-k_1^\mu)}{\ell^2\, s_{2,\ell}\cdots s_{23\cdots ,n,\ell}}
 +\cdots +
 \frac{N(\ell^\mu-k_{12\cdots ,n-1}^\mu)}{\ell^2\, s_{n,\ell}\cdots s_{n1\cdots ,n-2,\ell}}
\,,
\ea
it is easy to find the proper combination, {\it i.e.} 
rank-1 (vectorial) PT factors,
\ba
\label{ptvector}
{\rm PT}^{\mu}(1,2,\cdots,n):=  \sum_{i=1}^n (\ell^\mu- k_{12\cdots i-1}^\mu){\rm PT}^{\rm tree}(+,i,i+1,\cdots, i-1,-)\,,
\ea
for instance for $n=3$ we have
\ba
&{\rm PT}^{\mu}(1,2,3)
\\ \nonumber=
&\ell^\mu {\rm PT}^{\rm tree}(+,1,2,3,-)+(\ell^\mu-k_1^\mu) {\rm PT}^{\rm tree}(+,2,3,1,-)+(\ell^\mu-k_{12}^\mu) {\rm PT}^{\rm tree}(+,3,1,2,-)  \,,
\ea
such that
its one-loop CHY integral with the most symmetric factor gives an $n$-gon  with a non-trivial numerator
\ba
\frac{1}{\ell^2}\int d\mu_n  {\rm PT}^\mu(1,2,\cdots,n) \frac{(-1)^{n+1}}{\s_1\s_2\cdots\s_n} =\,\,&
\sum_{i=1}^n \frac{\ell^\mu-k_{12\cdots ,i-1}^\mu}{\ell^2\, s_{i,\ell}s_{i,i+1,\ell}\cdots s_{i,i+1,\cdots ,i-2,\ell}}
\\
\cong \,\,&
\ell^\mu  \,{\rm gon}(1,2,\cdots,n)
\non
\,.
\ea
Note that at the one-loop level, there is a global choice of loop momentum $\ell$ in the trivalent graphs. Inspired by this, the definition 
in \eqref{ptvector}  chooses $\ell$ is to be adjacent to leg $1$  on the left. With the same choice, we define higher-rank (tensorial) PT factors as in \eqref{ptvectorg}, and one can check that with the most symmetric factor we have 
\ba
\frac{1}{\ell^2}\int d\mu_n
{\rm PT}^{\mu_1,\mu_2,\cdots,\mu_r}(1,2,\cdots,n)
\frac{(-1)^{n+1}}{\s_1\s_2\cdots\s_n}
\cong \,\,&
\ell^{\mu_1}  \ell^{\mu_2} \cdots \ell^{\mu_r}    \,{\rm gon}(1,2,\cdots,n)
\,.
\ea
With this intuitive way to introduce the generalized PT factors two immediate questions are whether the CHY integrals of themselves would still be quadratic and how useful their integrals could be. We found the positive answer and their integrals act as the generalization of one-loop doubly ordered biadjoint amplitudes $m(\alpha |\beta)$.


Let us explain the basic properties of generalized PT factors themselves before turning to their CHY integrals.

\subsection{Properties of generalized PT factors}

Because of the relative positions of $\pm$ in the cyclic sum \eqref{ptvectorg},  none of the generalized PT factors have reverse symmetry,
\ba
\label{reverse}
{\rm PT}^{\mu_1,\mu_2,\cdots,\mu_r}(1,2,\cdots,n) \neq
{\rm PT}^{\mu_1,\mu_2,\cdots,\mu_r}(n,\cdots,2,1) ,\qquad
{\rm for} ~ r\geq 0.
\ea
Furthermore,  except for the original scalar PT factor,  they don't have cyclic symmetry, either.
However, inspired by the fact
\ba
\ell^{\mu_1}  \ell^{\mu_2} \cdots \ell^{\mu_r}    \,{\rm gon}(2,\cdots,n,1)  \cong \,{\rm gon}(1,2,\cdots,n)  \prod_{j=1}^r (\ell^{\mu_j}+k_1^{\mu_j})   \,,
\ea
one can easily check that the cyclic symmetry can be recovered up to some lower-rank PT factors
on the support of  MCEL, 
\ba
\label{pttenshift}
{\rm PT}^{\mu_1,\mu_2,\cdots,\mu_r}(2,\cdots,n,1)
=
\left( \prod_{j=1}^r ({ L}_1^{\mu_j}+k_1^{\mu_j})
\right)
{\rm PT}(1,2,\cdots,n)   \,,
\ea
where we have defined an operation
\footnote{
Note that  $L_1^\nu {\rm PT}^{\mu}(2,\cdots,n,1)
=
L_1^\nu \Big(  {\rm PT}^{\mu}(1,2,\cdots,n)+k_1^{\mu}  {\rm PT}(1,2,\cdots,n)\Big)
=
{\rm PT}^{\mu\nu}(1,2,\cdots,n)+k_1^{\mu} {\rm PT}^\nu(1,2,\cdots,n)
 $.}
\ba
\label{operation}
{ L}_1^{\mu} {\rm PT}^{\mu_1,\mu_2,\cdots,\mu_t}(1,2,\cdots,n)  =   {\rm PT}^{\mu_1,\mu_2,\cdots,\mu_t, \mu}(1,2,\cdots,n) \,,
\ea
and used the fact that the Lorentz indices in the generalized PT factors are totally symmetric. 
The generalization  $ { L}_1^{\mu_1\cdots \mu_r} = L_1^{\mu_1} \cdots L_1^{\mu_r}  $ is straightforward. Here is an
 example of \eqref{pttenshift}  (under the convention  $a^{(\mu} b^{\nu)}=a^{\mu} b^{\nu}+b^{\mu} a^{\nu}$) 
\ba
\label{pttenshiftex}
{\rm PT}^{\mu\nu}(2,\cdots,n,1)
=
{\rm PT}^{\mu\nu}(1,2,\cdots,n)+k_1^{(\mu} {\rm PT}^{\nu)}(1,2,\cdots,n)
+k_1^\mu k_1^\nu {\rm PT}(1,2,\cdots,n) \,.
\ea
So we see up to rank-$r$, there are $(r+1)(n-1)!$ algebraically linearly independent generalized PT factors.

Now we consider whether there are other relations among these $(r+1)(n-1)!$  generalized PT factors on the support of SE  in \eqref{oneloopchy}.  Since $\ell$ is sensitive when we apply the formula \eqref{numeratorngon} to construct quadratic propagators, naturally we require that all coefficients of generalized PT factors in the possible new relations should be free of  $\ell$. There are indeed such relations iff we contract some indices of the generalized PT factors  ${\rm PT}^{\mu \cdots}$ with external momentum vectors $k_{i,\mu}$, for example,
\ba
0 \cong& \sum_{\rho\in S_{n-1}}k_{1,\mu} {\rm PT}^\mu (1,\rho(2,\cdots,n)) \,,
\\
\non
0 \cong&
k_{23,\mu}\big(
{\rm PT}^\mu (1,2,3,4)
+{\rm PT}^\mu (1,3,2,4)
\big)
+
k_{2,\mu}{\rm PT}^\mu (1,3,4,2) 
+
k_{3,\mu}
{\rm PT}^\mu (1,2,4,3)\,.
\ea
So the rank of the generalized PT factors must be no more than $(r+1)(n-1)!$.  By construction, a single higher-rank PT factor cannot be reduced as a linear combination of lower-rank ones. Besides,
it's easy to check that the $(n-1)!$ scalar PT factors themselves are independent. Hence the rank is
no less than  $(n-1)!+r$. We leave the exploration of the exact rank for future work.

In addition to CHY integral with the most symmetric factor, in the following two sections, we will discuss the one-loop CHY integral of two scalar PT factors and then two generalized ones.




\section{Revisiting CHY integrals of scalar PT factors
\label{sec:inte1} }

In this section, we organize the results of CHY integrals of two scalar PT factors \eqref{onelooppt}  in a  way different from \cite{He:2015yua}, with a strict proof for all multiplicities put in the appendix \ref{proofpro}. 
Although the scalar cases are simple, we find them useful for introducing our notation and in particular the so-called {\it cyclic partitions}, 
using which the CHY integral of two arbitrary generalized PT factors 
 can be present in a way very similar to that of scalar ones.

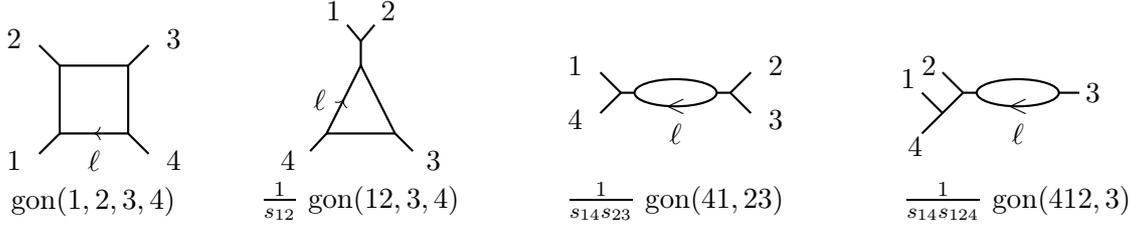
\begin{figure}
    \centering
\begin{tikzpicture}[decoration={
    markings,
    mark=at position 0.5 with {\arrow{>}}},scale=1.8]
\draw[thick](0,0)--(0.5,0)--(0.5,0.5)--(0,0.5)--(0,0)--(-0.15,-0.15); 
\draw[thick](0,0.5)--(-0.15,0.65);
\draw[thick](0.5,0.5)--(0.65,0.65);
\draw[thick](0.5,0)--(0.65,-0.15);
\draw[thick](-0.2,-0.2) node[left=0pt]{$1$};
\draw[thick](-0.2,0.7) node[left=0pt]{$2$};
\draw[thick](0.7,0.7) node[right=0pt]{$3$};
\draw[thick](0.7,-0.2) node[right=0pt]{$4$};
\draw[postaction={decorate}](0.5,0)--(0,0);
\draw (0.25,-0.2) node{$\ell$};
\draw[thick](0.25,-0.5) node{ gon$(1,2,3,4)$};

\draw[thick] (4.5,0.3) ellipse (0.3 and 0.1 );
\draw (4.5,0.2)node{\small $<$};
\draw(4.5,0)node{$\ell$};
\draw[thick](4.8,0.3)--(4.9,0.3);
\draw[thick](4.9,0.3)--(5.05,0.45);
\draw[thick](4.9,0.3)--(5.05,0.15);
\draw[thick] (5.1,0.5) node[right=0pt]{$2$};
\draw[thick] (5.1,0.1) node[right=0pt]{$3$};
\draw [thick](4.1,0.3)--(3.95,0.45);
\draw[thick] (3.9,0.5) node[left=0pt]{$1$};
\draw[thick] (4.1,.3)--(3.95,0.15);
\draw [thick](3.9,0.1) node[left=0pt]{$ 4$};
\draw[thick](4.2,0.3)--(4.1,0.3);
\draw[thick](4.5,-0.5) node{ $\frac{1}{s_{14}s_{23}}$ gon$(41,23)$};

\draw[thick](1.95,0)--(2.45,0)--(2.2,0.5)--(1.95,0)--(1.83,-0.13);
\draw[postaction={decorate}](1.95,0)--(2.2,0.5);
\draw (1.9,0.25)node{$\ell$};
\draw[thick](2.45,0)--(2.58,-0.13);
\draw[thick](1.82,-0.2) node[left=1pt]{$4$};
\draw[thick](2.58,-0.2) node [right=1pt]{$3$};
\draw[thick](2.2,0.5)--(2.2,0.68)--(2.1,0.8);
\draw[thick](2,0.9) node{ $1$};
\draw[thick](2.4,0.9) node{ $2$};
\draw[thick](2.2,0.68)--(2.3,0.8);
\draw[thick](2.2,-0.5) node{ $\frac{1}{s_{12}}$ gon$(12,3,4)$};

\draw[thick] (7,0.3) ellipse (0.3 and 0.1 );
\draw (7,0.2)node{\small $<$};
\draw(7,0)node{$\ell$};
\draw[thick](6.7,0.3)--(6.6,0.3);
\draw[thick](6.6,0.3)--(6.45,0.45);
\draw[thick](6.6,0.3)--(6.45,0.15)--(6.3,0);
\draw[thick](6.45,0.15)--(6.3,0.3);
\draw[thick] (6.35,0.5) node{$2$};
\draw[thick] (6.25,-0.1) node{$4$};
\draw[thick] (6.2,0.4) node{$1$};
\draw[thick] (7.3,0.3)--(7.45,0.3);
\draw[thick] (7.55,0.3) node{$3$};
\draw[thick](7,-0.5)node{ $\frac{1}{s_{14}s_{124}}$ gon$(412,3)$};
\end{tikzpicture}
    \caption{Some trivalent graphs produced by 
    $\frac{1}{\ell^2}\int d\mu_4  \Big( {\rm PT}(1,2,3,4)  \Big)^2 $
    }
    \label{fig:sometriva}
\end{figure}

Just \,like\, the\, tree\, level, the one-loop\, CHY\, integral\, of\, two\, identical\, PT\, factors ${\rm PT}(1,\rho(2,\cdots,n))^2$ produces all one-loop trivalent graphs with external legs consistent with the ordering $(1,\rho(2,\cdots,n))$ except the tadpoles. Here are a few simple examples
\ba
\label{s123s123}
\frac{1}{\ell^2}\int d\mu_3  \Big( {\rm PT}(1,2,3)  \Big)^2  \cong \,\,&
{\rm gon}(1,2,3)+ \frac{{\rm gon}(12,3) }{s_{12}}+ \frac{{\rm gon}(1,23) }{s_{23}}+ \frac{{\rm gon}(31,2) }{s_{13}}
\,,
\\
\label{s1234s1234}
\frac{1}{\ell^2}\int d\mu_4  \Big( {\rm PT}(1,2,3,4)  \Big)^2  \cong \,\,&
{\rm gon}(1,2,3,4)+ \frac{{\rm gon}(12,34) }{s_{12}s_{34}}+ \frac{{\rm gon}(41,23) }{s_{41}s_{23}}
\\
\non
\,\,&+
\Bigg[\Bigg( \frac{{\rm gon}(12,3,4) }{s_{12}}+ \frac{{\rm gon}(123,4) }{s_{123}}(\frac{1}{s_{12}}+\frac{1}{s_{23}})
\Bigg)+ {\rm cyc}(1234)\Bigg]
\,,
\ea
with some  trivalent graphs shown in 
figure \ref{fig:sometriva}.
%
To have a better description of all one-loop trivalent graphs consistent with a particular ordering, we find it useful to introduce cyclic partitions to describe both the polygons and sub-trees.


{\textbf {Definition:}} A {\it cyclic partition} means to cut a cyclic ordering $(1,\rho(2,\cdots,n))$ into several non-empty sequences where the one containing the leg $1$ is denoted as $A_1=A1{\bar A}$ and the following as $A_2,A_3,\cdots, A_m$. Besides, we denote the sub-sequences before and after 1 in $A_1$ as $A$ and  ${\bar A}$, which could be empty.  Obviously, we have $1\leq m\leq n$.

For example, for the ordering $(12)$, we have 3 cyclic partitions,
\ba
\{A_1=1, A_2=2\},\quad \{A_1=12\},\quad \{A_1=21\}\,.
\ea
 For the ordering
$(123)$, we have 7 cyclic partitions,
\ba
&\{A_1=1, A_2=2, A_3=3 \},\quad \{A_1=12, A_2=3 \},\quad \{A_1=31, A_2=2 \},\quad \{A_1=1, A_2=23 \},
\nl
&\{A_1=123\},\quad \{A_1=312\},\quad \{A_1=231\}\,.
\ea
It's easy to see for cyclic partition with exactly $m$ sequences there are ${n \choose m}$  different partitions and hence $\sum_{m=1}^n {n \choose m} =2^n-1$   in total.  For example, the  ${4 \choose 2}=6$ $2$-sequence partitions for the ordering  $(1243)$ are given by
\ba
&\{A_1=124, A_2=3\},\quad \{A_1=12, A_2=43\},
\quad \{A_1=1, A_2=243\},
\nl
&\{A_1=431, A_2=2\},\quad \{A_1=31, A_2=24\},
\quad \{A_1=312, A_2=4\},
\ea
where for the first cyclic partition we  have $A=\emptyset$ and $\bar A=24$ while for the last one, $A=3$ and $\bar A=2$.

\subsection{Two identical scalar  PT factors}

An particular $m$-sequence partition, $\{A_1, \cdots, A_m\}$ can immediately be used to construct an particular $m$-gon, i.e.,  ${\rm gon}(A_1,\cdots, A_m)$. All sub-trees planted on every corner of the polygons that are consistent with the ordering $(1,\rho(2,\cdots,n))$ can be described by the
effective Feynman diagrams \cite{Cachazo:2013iea,Huang:2018zfi,Feng:2019xiq,Feng:2020opo}, or 
 the equivalent Berends-Giele double-current
of bi-adjoint $\phi^3$ theories with identical ordering
$\phi_{A_i\mid\tilde{A}_i}$
\cite{Mafra:2016ltu}. For example,
\ba
&\phi_{1\mid 1}=1,\quad
\phi_{12\mid 12}=\frac{1}{s_{12}}
,\quad
\phi_{123\mid 123}=\frac{1}{s_{123}}\left( \frac{1}{s_{12}}+\frac{1}{s_{23}}\right)\,.
\ea
Roughly speaking, $\phi_{12\dots r\mid 12\dots r}$  is nothing but an overall pole $\frac{1}{s_{12\cdots r}}$ times a $(r+1)$-point $\phi^3$ tree-level amplitude $m(12\cdots r+1|12\cdots r+1  )$ without using the momentum $k_{r+1}$.

Then all one-loop trivalent graphs except the tadpoles corresponding to the ordering $(1,\rho(2,\cdots,n))$ can be expressed as
\ba
\label{scalarid}
\frac{1}{\ell^2}\int d\mu_n  \Big( {\rm PT}(1,\rho(2,\cdots,n))  \Big)^2 \cong
\!\!\!
\sum_{
\substack
{
(A_1,A_2,\cdots, A_m)=(1,\rho)
\\
2\leq m\leq n}
}
\!\!\!
{\rm gon}(A_1,A_2, \cdots, A_m) \prod_{i=1}^m  \phi_{A_i\mid A_i}\,,
\ea
which are proved to give the one-loop CHY integral of two identical scalar PT factors in appendix \ref{proofpro}.

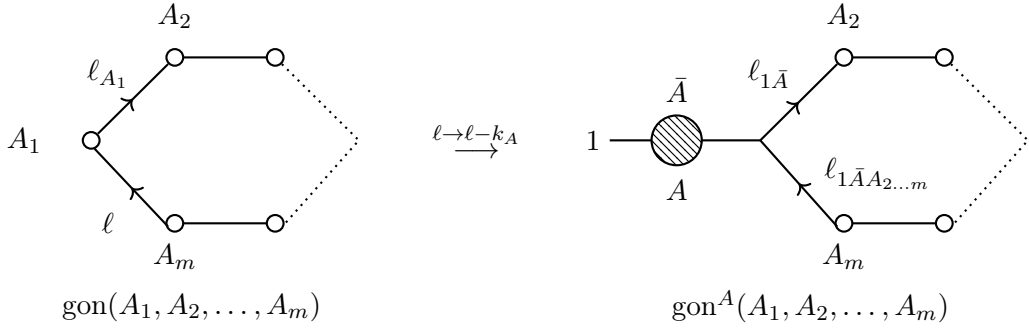
\begin{figure}
    \centering
\begin{tikzpicture}[decoration={
    markings,
    mark=at position 0.5 with {\arrow{>}}},
place/.style={circle,draw,fill=gray,
inner sep=0pt,minimum size=1mm},scale=2.2]
\draw[thick] (1,1) circle (0.05);
\draw[thick](0.6,1) node { $A_1$};
\draw [thick](1.5,1.5) circle (0.05);
\draw[thick] (1.5,1.75) node{$A_2$};
\draw [thick](1.5,0.5) circle (0.05);
\draw[thick] (1.5,0.3) node{ $A_m$};
\draw[postaction={decorate},thick] (1.04,1.04)--(1.46,1.46);
\draw[thick] (1.1,1.4) node{$\ell_{A_1}$};
\draw[thick] [postaction={decorate}](1.46,0.48)--(1.02,0.95);
\draw[thick](1.1,0.5)node{ $\ell$};
\draw[thick](2.1,1.5) circle(0.05);
\draw[thick](2.1,0.5) circle(0.05);
\draw[thick](1.55,0.5)--(2.05,0.5);
\draw[thick](1.55,1.5)--(2.05,1.5);
\draw[thick][dotted](2.14,1.46)--(2.6,1)--(2.16,0.5);
\draw[thick](1.6,0) node{ gon$(A_1,A_2,\dots,A_m)$};

\draw[thick](3.3,1) node{$\overset{\ell\rightarrow \ell{-}k_{A}}{\longrightarrow}$};

\draw[thick] (5.5,1.5) circle (0.05);
\draw[thick] (5.5,1.75) node{ $A_2$};
\draw[thick] (5.5,0.5) circle (0.05);
\draw[thick] (5.5,0.3) node{ $A_m$};
\draw[thick][postaction={decorate}] (5,1)--(5.46,1.46);
\draw[thick] (5.05,1.4) node{$\ell_{1\bar{A}}$};
\draw[thick] [postaction={decorate}](5.46,0.48)--(5,1);
\draw[thick](5.7,0.8)node{ $\ell_{1\bar{A}A_{2\dots m}}$};
\draw[thick](6.1,1.5) circle(0.05);
\draw[thick](6.1,0.5) circle(0.05);
\draw[thick](5.55,0.5)--(6.05,0.5);
\draw[thick](5.55,1.5)--(6.05,1.5);
\draw[thick][dotted](6.14,1.46)--(6.6,1)--(6.16,0.5);
\draw[thick](5,1)--(4.65,1);
\draw[thick][pattern=north west lines] (4.5,1) circle(0.15);
\draw[thick](4.35,1)--(4.1,1);
\draw[thick](4,1) node {$1$};
\draw[thick](4.5,1.3) node{$\bar{A}$};
\draw[thick](4.5,0.7) node{$A$};
\draw[thick](5.3,0)node{ gon$^A(A_1,A_2,\dots,A_m)$};
\end{tikzpicture}
    \caption{Shift $\ell$ to make it adjacent to leg 1  on the left.
    Each corner may contain multi external legs forming a trivalent sub-tree planted on the polygons. The corner $A_1$ is divided into three parts $A, 1 $ and ${\bar A}$ where $A$ and  ${\bar A}$ could be empty.
    }
    \label{fig:shift}
\end{figure}

On the right hand side of the above equation, there is an overall loop propagator $\frac{1}{\ell^2}$ and the loop momentum $\ell$ is put just after the first corner $A_1$ in every one-loop trivalent graph.   To have a universal position of loop  momentum
$\ell$ for every  graph, as shown in figure \ref{fig:shift}, we can make use of the shift $\ell\to \ell-k_{A}$,
\ba
{\rm gon}^A(A1{\bar A}, A_2,\cdots, A_m):={\rm gon}(A1{\bar A}, A_2,\cdots, A_m)\big|_{\ell \to  \ell-k_{A}}=\,\,& \frac{1}{ \ell_{1\bar A}^2 \ell_{1{\bar A},A_{2}}^2 \cdots \ell_{1{\bar A},A_{23\cdots ,m} }^2}\,,
\ea
and rewrite the RHS of \eqref{scalarid} as
\ba
\label{defgs}
G(1,\rho(2,\cdots,n)):=\sum_{
\substack
{
(A_1,\cdots, A_m)=(1,\rho)
\\
2\leq m\leq n}
}
\!\!\!
{\rm gon}^A(A_1, \cdots, A_m) \prod_{i=1}^m  \phi_{A_i\mid A_i}\,,
\ea
which we define as $G$. In this representation, the loop momentum is always connected to leg 1 on the left in every graph.  For example,
\ba
\label{s123s123}
\frac{1}{\ell^2}\int d\mu_3  \Big( {\rm PT}(1,2,3)  \Big)^2  \cong \,\,G(123)=
{\rm gon}(1,2,3)+ \frac{{\rm gon}(12,3) }{s_{12}}+ \frac{{\rm gon}(1,23) }{s_{23}}+ \frac{{\rm gon}^3(31,2) }{s_{13}}
\,.
\ea

\subsection{Two different scalar PT factors \label{diorder}}

Just like at the tree level, the one-loop CHY integral of two different scalar  PT factors should correspond to all one-loop trivalent graphs except tadpoles that are consistent with both orderings, i.e., it can be obtained by the intersection of those of identical scalar  PT factors. Formally, we can write
\ba
\label{scalardi}
&\frac{1}{\ell^2}\int d\mu_n  {\rm PT}(1,\rho(2,\cdots,n))
{\rm PT}(1,\sigma(2,\cdots,n))
 \cong
\!\!\!
\sum_{
\substack
{
(A_1,\cdots, A_m)=(1,\rho)
\\
({\tilde A}_1,\cdots, {\tilde A}_m)=(1,\sigma)
\\
\{A_j\}=\{{\tilde A}_j\}
\\
2\leq m\leq n}
}
\!\!\!
{\rm gon}(A_1, \cdots, A_m) \prod_{i=1}^m  \phi_{A_i\mid\tilde{A}_i}\,,
\non
\\[-10mm]
\ea
where we have  used  $\{A_j\}= \{\tilde A_j\}$ to emphasize that
the sequences $A_j$ and $\tilde A_j$ are  the same up to a
permutation of their elements for all $j=1,...,m$.
$\phi_{A_i\mid\tilde{A}_i}$ is again 
the Berends-Giele double-current
of bi-adjoint $\phi^3$ theories with two general orderings
which can be  obtained by
\ba
\phi_{A_i\mid\tilde{A}_i}= {\rm sgn} ^{A_i}_{\tilde A_i} \, \phi_{A_i\mid A_i}\bigcap \phi_{\tilde{A}_i\mid \tilde{A}_i}\,,
\ea
where the sign  is defined as
\ba
\sgn^{\a}_\b:=
\prod_{i=1}^{|\beta|-1}\sgn^{\a}_{\b(i),\b(i+1)} = \sgn^{\b}_\a
\qquad
{\rm with}
~
\label{defsign}
\sgn^{\a}_{i,j}:=
\begin{cases}
+1 \quad {\text {if~$i$ precedes $j$ in $\a$}}\,,
\\
-1 \quad {\text {if~$j$ precedes $i$ in $\a$}}\,.
\end{cases}
\ea
For example, $\sgn^{234}_{23}=\sgn^{234}_{24}=\sgn^{234}_{34}=\sgn^{234}_{234}=1, \sgn^{234}_{32}=\sgn^{234}_{43}=\sgn^{234}_{243}=-1 $.
Here are examples of $\phi$ with different orderings,
\ba
\phi_{23\mid 32}&=-\frac{1}{s_{23}}\,,
\quad
\phi_{234\mid243}=-\frac{1}{s_{234} s_{34}}\,,
\quad
\phi_{2345\mid 3254}=\frac{1}{s_{2345} s_{23}s_{45}}.
\ea
By simple calculation, one can show that
the sign of each term in the summation of \eqref{scalardi}  actually becomes an overall sign given by
${\rm sgn}^{\rho}_{\sigma} $.

Recall that because of the relative positions of $\pm$, none of the PT factors have reverse symmetry \eqref{reverse}. Correspondingly, in the trivalent diagrams  produced by CHY integrals,  even though the sub-trees don't care about the 
relative positions of $\pm \ell$, the polygons \eqref{partialgeneral} do care about the orientations of the loop momenta. 
Hence the CHY integral of ${\rm PT}(1,\rho)$ and ${\rm PT}(1,\sigma)$ won't equal to that of ${\rm PT}(1,\rho^{T})$ and  ${\rm PT}(1,\sigma)$ in general. In particular, the  integral of ${\rm PT}(1,\rho)$ and its reverse ${\rm PT}(1,\rho^{T})$ won't produce any $n$-gon as long as $n\geq 3$, which is obviously different from that of two identical  ${\rm PT}(1,\rho)$ \eqref{scalarid}. Actually, two reverse orderings can only produce various bubbles.  For example,
\ba
 \frac{1}{\ell^2}\int d\mu_4   {\rm PT}(1,2,3,4)
{\rm PT}(1,4,3,2) \cong \,\,&
\frac{{\rm gon}(12,34) }{s_{12}s_{34}}+ \frac{{\rm gon}(41,23) }{s_{41}s_{23}}
\label{s1234s4321}
\nl
\,\,&+
\Bigg[ \frac{{\rm gon}(123,4) }{s_{123}}(\frac{1}{s_{12}}+\frac{1}{s_{23}})
+ {\rm cyc}(1234)\Bigg]\,.
 \ea
If we reverse two orderings at the same time, i.e., ${\rm PT}(1,\rho){\rm PT}(1,\sigma)$ vs. ${\rm PT}(1,\rho^{T}){\rm PT}(1,\sigma^{T})$, their integrals turn back to be the same up to an overall orientation of loop momentum $\ell \to -\ell$ as expected. For example,
\ba
 \frac{1}{\ell^2}\int d\mu_4   {\rm PT}(1,2,3,4)
{\rm PT}(1,2,4,3)  
  \cong &
 -\frac{{\rm gon}(1,2,34)}{s_{34}}-\frac{{\rm gon}(1,234)}{s_{34} s_{234}}-\frac{{\rm gon}(12,34)}{s_{12} s_{34}}
 \nl&
 -\frac{{\rm gon}(123,4)}{s_{12} s_{123}}-\frac{{\rm gon}(124,3)}{s_{12} s_{124}}-\frac{{\rm gon}(134,2)}{s_{34} s_{134}}
\,,
\label{s1234s1243}
\\
 \frac{1}{\ell^2}\int d\mu_4   {\rm PT}(1,4,3,2)
{\rm PT}(1,3,4,2)  
  \cong &
 -\frac{{\rm gon}(1,34,2)}{s_{34}}-\frac{{\rm gon}(1,234)}{s_{34} s_{234}}-\frac{{\rm gon}(12,34)}{s_{12} s_{34}}
 \nl&
 -\frac{{\rm gon}(123,4)}{s_{12} s_{123}}-\frac{{\rm gon}(124,3)}{s_{12} s_{124}}-\frac{{\rm gon}(134,2)}{s_{34} s_{134}}
\,,
\ea
where ${\rm gon}(1,2,34)\cong {\rm gon}(1,34,2)|_{\ell\to -\ell}$.

Similar to \eqref{defgs},  we can shift every loop momentum $\ell$ on the RHS of \eqref{scalardi} such that $\ell$ is always connected to leg 1 on the left. We define the new representation  as
\ba
\label{defgs2}
&G\big( 1,\rho(2,\cdots,n)|1,\sigma(2,\cdots,n))
:=
\,
{\rm sgn}^\rho_\sigma
\sum_{
\!\!\!\!\!\!
\substack
{
(A_1,\cdots, A_m)=(1,\rho)
\\
({\tilde A}_1,\cdots, {\tilde A}_m)=(1,\sigma)
\\
\{A_j\}=\{{\tilde A}_j\}
\\
2\leq m\leq n}
}
\!\!\!\!\!\!
{\rm gon}^A(A_1, \cdots, A_m) \prod_{i=1}^m \left[ \phi_{A_i\mid A_i}\bigcap \phi_{\tilde{A}_i\mid \tilde{A}_i} \right]\,.
\non
\\[-10mm]
\ea
Here are
 some examples,
 %
%
\ba
 G(123|132)
=&
- \frac{{\rm gon}(12,3) }{s_{12}}- \frac{{\rm gon}(1,23) }{s_{23}}- \frac{{\rm gon}^3(31,2) }{s_{13}}
\cong \frac{1}{\ell^2}\int d\mu_3   {\rm PT}(1,2,3)  {\rm PT}(1,3,2)
\,,
\non
\\
G(12345|15324)
  =&
\frac{\text{gon}(1,2345)}{s_{23}
   s_{234}s_{2345}}+\frac{\text{gon}^{45}(451,23)}{s_{15}s_{23}s_{145}}+\frac{\text{gon}^5(51,234)}
   {s_{23}s_{234}s_{15}}+\frac{\text{gon}^5(5123,4)}{s_{15}
   s_{23}s_{1235}}\nl &
   +\frac{\text{gon}^{45}(4512,3)}{s_{15}
   s_{145}s_{1245}}+\frac{\text{gon}^{345}(3451,2)}{s_{15}
   s_{145}s_{1345}}+\frac{\text{gon}(1234,5)}{s_{23}
   s_{234}s_{1234}}+\frac{\text{gon}^5(51,23,4)}{s_{23}s_{15}}\,.
\ea

We emphasize that the CHY integrals are just equivalent to a set of one-loop trivalent graphs (up to an overall sign) on the support of loop integral while $G$ functions are defined to be exactly this set of 
trivalent graphs.  Besides, even though we have $G(1,\rho| 1,\sigma )\cong  G(1,\sigma|1,\rho )$, they are not exactly the same at algebraic level in general, i.e., $G(1,\rho| 1,\sigma )\neq  G(1,\sigma|1,\rho ) $.  This is  because  we have just taken the shift $\ell\to \ell-k_A$ instead of $\ell\to \ell-k_{\tilde A}$ where the shifts are obtained  
by  matching $A_1=A1{\bar A}$ and ${\tilde A}_1={\tilde A}1{\tilde {\bar A}}$ respectively.  We make this distinguishing as we are going to dress these $G$ functions with $\ell$-dependent numerators which are sensitive to the shifts of loop momenta when we talk about the CHY integrals of generalized PT factors.

\subsection{Absence of tadpoles} \label{disatad}
In \eqref{scalardi}, it seems that all tadpole contributions are dropped off by hand as we
excluded the case $m=1$ in the summation.  Actually, we can include these contributions in the formula as well but they add up to zero by $U(1)$ identity. Parameterizing
\ba
&A_1=\rho(a+1,a+2,\cdots, a), \qquad \forall a\in \{1,2,\cdots,n\}\,,
\nl&
{\tilde A}_1=\sigma(b+1,b+2,\cdots, b),\qquad \forall b\in \{1,2,\cdots,n\}\,,
\ea
with  identification $\rho(1)=\sigma(1)=1$, the possible tadpole contributions are given by
$
\frac{1}{\ell^2}
 \phi_{A_1 |\tilde{A}_1}
$.
However, when we sum over  $a$,
\ba
\label{u1cancel}
\sum_{a=1}^n
\frac{1}{\ell^2}
 \phi_{\rho(a+1,a+2,\cdots, a)| {\tilde A}_1}=0\,,
\ea
or sum over $b$, all of the tadpoles cancel because of the $U(1)$ identity even without using the momentum conservation of external $n$ legs. 
This explains why there are no tadpoles in \eqref{scalardi} or \eqref{scalarid}.  Similar arguments are also given in \cite{He:2015yua} but here we derive the results without complicated analysis of combinatorial structures of trivalent graphs and our arguments hold for more general cases.
Actually, as long as there is one scalar PT factor (since we just need to sum over $a$ or $b$), there will still be no tadpoles, as we will see in the next section.

\section{CHY integrals with generalized PT factors
\label{sec:inte2} }

Now we move to the one-loop CHY integrals of generalized PT factors with tensor structures ${\rm PT}^{\mu_1...\mu_r}(1,\rho(2,\cdots,n))\times {\rm PT}^{\nu_1...\nu_t}(1,\sigma(2,\cdots,n))$, which will produce one-loop trivalent graphs with non-trivial numerators.
Let us take a look at some examples first
\ba
\label{s123123new}
\frac{1}{\ell^2}\int d\mu_3   {\rm PT}^\mu (1,2,3)   {\rm PT}(1,2,3)   \cong& \,\,
\ell^\mu \left[{\rm gon}(1,2,3)+ \frac{{\rm gon}(12,3) }{s_{12}}+ \frac{{\rm gon}(1,23) }{s_{23}}+ \frac{{\rm gon}^3(31,2) }{s_{13}}\right]\,,
\nl
\frac{1}{\ell^2}\int d\mu_3   {\rm PT}^{\mu\nu}(1,2,3)  {\rm PT}(1,3,2)   \cong &
-
\ell^\mu\ell^\nu\left[\frac{{\rm gon}(12,3) }{s_{12}}+ \frac{{\rm gon}(1,23) }{s_{23}}+ \frac{{\rm gon}^3(31,2) }{s_{13}}
\right],
\\
\frac{1}{\ell^2}\int d\mu_2   {\rm PT}^\mu (1,2)   {\rm PT}^\nu(1,2)  \cong&\,
\ell^\mu\ell^\nu {\rm gon}(1,2)+\frac{k_1^\mu k_1^\nu}{\ell^2s_{12}}\,,
\label{v12v12}
\\
\frac{1}{\ell^2}\int d\mu_3   {\rm PT}^\mu (1,2,3)   {\rm PT}^\nu(1,3,2)   \cong &\,\,
-\ell^\mu \left[ \frac{(\ell^\nu+k_2^\nu){\rm gon}(12,3) }{s_{12}}+ \frac{ \ell^\nu {\rm gon}(1,23) }{s_{23}}+ \frac{\ell^\nu {\rm gon}^3(31,2) }{s_{13}}\right]
\nl
&
-
\frac{1}{\ell^2s_{123}} \left(
\frac{k_1^\mu k_1^\nu}{s_{23}}
-\frac{k_2^\mu k_{13}^\nu}{s_{13}}
-\frac{k_{12}^\mu k_{3}^\nu}{s_{12}}
\right)\,.
\label{v123v132}
\ea
We see that there are no tadpole contributions as long as at least one of them is a scalar PT factor. Besides, loop momenta in the numerators are no longer overall factors in general. In the following, we give the proposal for the general integrals in an intuitive way,  specify possible simplifications for various cases and leave the strict proof in appendix \ref{proofpro}, including those of \eqref{scalarid} and \eqref{scalardi}.

\subsection{Intersections of two orderings}
As already discussed in section \ref{diorder},
nonzero contributions can only come from the intersection of two orderings. For example, for a given $m$-gon,  the sequences $A_j$ and $\tilde A_j$ are required to be the same up to a
permutation of their elements for all $j=1,...,m$.
The main challenge now is how to deal with the very nontrivial $\ell$-dependent numerators coming from both Parke-Taylor factors.

Naively, for the $m$-gon, the PT factor ${\rm PT}^{\mu_1...\mu_r}(1,\rho(2,\cdots,n))$ would give
\ba\label{eq54}
 \ell^{\mu_1}... \ell^{\mu_r} {\rm gon}^A(A_1, \cdots, A_m) \prod_{i=1}^m  \phi_{A_i\mid A_i}\,,
\ea
while the other ${\rm PT}^{\nu_1...\nu_t}(1,\sigma(2,\cdots,n))$ would give
\ba\label{eq55}
 {\tilde \ell}^{\nu_1}... {\tilde\ell}^{\nu_t} {\tilde {\rm gon}}^{\tilde A} ({\tilde A}_1, \cdots, {\tilde A}_m) \prod_{i=1}^m  \phi_{\tilde{A}_i\mid\tilde{A}_i}\,,
\ea
where ${\tilde {\rm gon}}^{\tilde A} ({\tilde A}_1, \cdots, {\tilde A}_m) = {\rm gon}^{\tilde A} ({\tilde A}_1, \cdots, {\tilde A}_m) |_{\ell\to {\tilde \ell}}$.
Reall that
\ba
{\rm gon}^A(A_1, \cdots, A_m)&=\frac{1}{\ell^2_{1{\bar A}}\ell^2_{1{\bar A}A_1},\cdots, \ell^2_{1{\bar A}A_{2\cdots m}} } ,
\nl
{\tilde {\rm gon}}^A({\tilde A}_1, \cdots, {\tilde A}_m)&=\frac{1}{{\tilde \ell}^2_{1{\tilde {\bar A}}}{\tilde \ell}^2_{1{\tilde {\bar A}}{\tilde A}_2},\cdots, {\tilde \ell}^2_{1{\tilde {\bar A}}{\tilde A}_{2\cdots m}} }
=
\frac{1}{{\tilde \ell}^2_{1{\tilde {\bar A}}}{\tilde \ell}^2_{1{\tilde {\bar A}}{ A}_2},\cdots, {\tilde \ell}^2_{1{\tilde {\bar A}}{ A}_{2\cdots m}} } \,,
\ea
where in the last equality, we have used  $k_{A_i}=k_{{\tilde A}_i}$.
In order to have an intersection, we have to identify these two polygons, {\it i.e.}, we require
\ba \tilde\ell+k_1+k_{\tilde {\bar A}}=\ell+k_1+k_{{\bar A}}\,,
~~~\label{rule-5}\,
\ea
which means that their loop momentum is related as
\ba
 \tilde\ell= \ell-\Delta_{A{\tilde A} }, \qquad {\rm with}~~~~\Delta_{A{\tilde A} }=k_{\tilde {\bar A}}-
k_{\bar A}=
k_{ A}-
k_{\tilde { A}}\,.
~~~\label{rule-6}
\ea
In this way we find the final contribution to this $m$-gon from the pair:
\ba
 \ell^{\mu_1}... \ell^{\mu_r}  (\ell-\Delta_{A{\tilde A} })^{\nu_1}...(\ell-\Delta_{A{\tilde A} })^{\nu_t}\, {\rm gon}^A(A_1, \cdots, A_m) \prod_{i=1}^m  \phi_{A_i\mid\tilde{A}_i}\,,
\ea
where
the numerator $\ell^{\mu_1} \cdots \ell^{\mu_r}$ in \eqref{eq54} and that in \eqref{eq55} are just  multiplied to the $m$-gon with possible sub-trees planted on it.

\subsection{A general formula with two generalized PT factors}
Considering all polygons including tadpoles, we propose the following formula for two  generalized PT factors, which is the main result of our paper:
\ba
\label{twogeneral}
&\frac{1}{\ell^2}\int d\mu_n
{\rm PT}^{\mu_1,\cdots,\mu_r}
(1,\rho(2,\cdots,n))
 {\rm PT}^{\nu_1,\cdots,\nu_t}
(1,\sigma(2,\cdots,n))
 \\
 \non
\cong&
 \ell^{\mu_1} \cdots
\ell^{\mu_r}
\!\!\!\!\!\!\!\!\!\!\!\!
\sum_{
\substack
{
(A_1,A_2,\cdots,A_m)=(1,\rho)\\
(\tilde A_1, \tilde A_2,\cdots, \tilde A_m)=(1,\sigma)
\\
\{A_j\}= \{\tilde A_j\}
\\
1\leq m\leq n
}}
\!\!\!\!\!\!\!\!\!\!\!\!
(\ell-\Delta_{A{\tilde A} })^{\nu_1}...
(\ell-\Delta_{A {\tilde A} })^{\nu_t}
{\rm gon}^A(A_1, \cdots, A_m) \prod_{i=1}^m  \phi_{A_i\mid\tilde{A}_i}\,,
\ea
where  $A_1= A1{\bar A}$,
${\tilde A}_1={\tilde  A}1{\tilde {\bar A}}$ and $\Delta_{A{\tilde A} }=
k_{ A}-
k_{\tilde { A}}$.
We see that $\ell^\mu$'s are overall factors but  $\ell^\nu$'s are not. They cannot be made as overall factors at the same time in general.  Except the tadpole cases $m=1$, the sign coming from $\prod \phi$ actually becomes an overall sign given by ${\rm sgn}^\rho_\sigma$.
   Here we list two more examples in addition to \eqref{s123123new}-\eqref{v123v132},
\ba
\label{vv123v132}
&\frac{1}{\ell^2}\int d\mu_3   {\rm PT}^{\mu\gamma} (1,2,3)   {\rm PT}^\nu(1,3,2)
\\
 \cong & \,\,
-\ell^\mu \ell^\gamma \left[ \frac{(\ell^\nu+k_2^\nu){\rm gon}(12,3) }{s_{12}}+ \frac{ \ell^\nu {\rm gon}(1,23) }{s_{23}}+ \frac{\ell^\nu {\rm gon}^3(31,2) }{s_{13}}\right]
+{\rm tadpoles}\,,
\nl
&\frac{1}{\ell^2}\int d\mu_4   {\rm PT}^\mu (1,2,3,4)   {\rm PT}^\nu (1,3,2,4)
\label{v1234v1324}
\\
  \cong&
-\ell^\mu
\Bigg[ \text{gon}^4(41,23) \ell^{\nu
   }
    +\text{gon}(1,23,4)
   \ell^{\nu }
   +\frac{\text{gon}(1,234)  \ell^{\nu
   }}{s_{23}}+\frac{\text{gon}(123,4) \ell^{\nu }}{s_{23}}
   \nl&
  \qquad\,
   +\frac{\text{gon}^4(412,3)  \left(\ell^{\nu }+k_2^{\nu }\right)}{s_{14}}+\frac{\text{gon}^{34}(341,2) \left(\ell^{\nu }-k_3^{\nu }\right)}{s_{14}}
   \Bigg]+{\rm tadpoles}\,,
   \non
  \ea
where the tadpole parts will be completed soon in  \eqref{vv123v132com} and \eqref{v1234v1324com}.
Let's take the last line of the above equation as an example to explain the shift $\Delta_{A,\tilde A}$. For $\text{gon}^4(412,3) $, the corner containing particle 1 consists of particles $\{1,2,4\}$. In the ordering of the first PT factor, it gives $A_1=412$, i.e., $A=4$ while in the second ordering, it gives ${\tilde A}_1=241$, i.e., ${\tilde A}=24$. Together, they give rise to  $\Delta_{A,\tilde A}=k_{4}-k_{24}=-k_2 $ hence we have $\ell-\Delta_{A,\tilde A}=\ell+k_2$ in the numerator.  Similarly, for $\text{gon}^{34}(341,2)$, we have $A_1=341$ and ${\tilde A}_1=413$, which leads to $\Delta_{A,\tilde A}=k_{34}-k_{4}=k_3 $.

There are some simplifications for certain situations or certain terms in \eqref{twogeneral},
 which we specify next.  The proof of the proposal is put in appendix \ref{proofpro}.
Since all generalized PT factors can be reduced to those with 1 as their first entries because of \eqref{pttenshift}, the one-loop CHY integrals with arbitrary orderings can always be reduced as a combination of \eqref{twogeneral}.
In appendix \ref{twomostgen}, we present a formula to give their integrals directly.

\subsubsection{With one scalar PT factor}
When there are no $\nu$ indices in \eqref{twogeneral}, obviously all loop momentum in the numerators become overall factors and
the  one-loop CHY integral becomes strikingly simple
\ba
\label{onegeneral}
&\frac{1}{\ell^2}\!\int d\mu_n
{\rm PT}^{\mu_1,\cdots,\mu_r}
(1,\rho(2,\cdots,n))
{\rm PT}(1,\sigma(2,\cdots,n))
\cong
 \ell^{\mu_1} \cdots
\ell^{\mu_r}
G(1,\rho(2,..,n) | 1,\sigma(2,...,n))\,.
\ea
See some examples in \eqref{s123123new}.
The key point is that we have insisted on a universal position of loop momentum $\ell$ in the definition of $G$ functions \eqref{defgs2} which makes the overall factors possible.
Because of  the existence of the scalar PT factor, ${\rm PT}(1,\sigma(2),...,\sigma(n))$, there is no tadpole
topology based on an analysis analog to section \ref{disatad}.

\subsubsection{Identical ordering}

If $\rho=\sigma$, all $\Delta_{A \tilde A}=0$ for $m$-gons with $m\geq 2$. But for the
$1$-gons (tadpoles), $\Delta_{A \tilde A}\neq 0$ by different choices of the last elements of $A_1$ and $\tilde{A}_1$ in \eqref{twogeneral}.
So as a special case of \eqref{twogeneral},  the integral of two generalized PT factors with identical ordering is given by
\ba
\label{twogeneralid}
\frac{1}{\ell^2}\int d\mu_n
{\rm PT}^{\mu_1,\cdots,\mu_r}
(1,\rho(2,\cdots,n))
 {\rm PT}^{\nu_1,\cdots,\nu_t}
(1,\rho(2,\cdots,n))
\cong &
\,
 \ell^{\mu_1} \cdots
\ell^{\mu_r}
 \ell^{\nu_1} \cdots
\ell^{\nu_t}
G(1,\rho(2,..,n))
\nl
&+{\rm tadpoles}\,,
\ea
where the tadpole part will be discussed immediately.
For example,
\ba
\label{v123v123}
\frac{1}{\ell^2}\int d\mu_3   {\rm PT}^\mu (1,2,3)   {\rm PT}^\nu(1,2,3)   \cong \,\,&
\ell^\mu\ell^\nu \left[{\rm gon}(1,2,3)+ \frac{{\rm gon}(12,3) }{s_{12}}+ \frac{{\rm gon}(1,23) }{s_{23}}+ \frac{{\rm gon}^3(31,2) }{s_{13}}\right]
\nl
&
+
\frac{1}{\ell^2s_{123}} \left(
\frac{k_1^\mu k_1^\nu}{s_{23}}
+\frac{k_2^\mu k_2^\nu}{s_{13}}
+\frac{k_{12}^\mu k_{12}^\nu}{s_{12}}
\right)\,.
\ea

\subsection{Tadpoles \label{sectiontadpole}}
Since each generalized PT factor consists of $n$ tree-level PT factors  \eqref{ptvectorg} where the legs $+$ and $-$ are adjacent, the $(n+2)$-pt tree-level CHY integral of any of these two tree-level PT factors would  produce some tree-level trivalent graphs where the  legs $+$ and $-$
meet at a vertex.  In the forward limit, all of these
 graphs would contribute to tadpoles \eqref{oneloopchyori}.  Considering all of $n^2$ tree-level integrals   when we expand the one-loop CHY integral in \eqref{twogeneral}, we get the following  tadpole results
 \footnote{Here 
 $k_{1\rho({23\cdots a} )}$ comes from the coefficients in \eqref{ptvectorg}.
 So when $a=n$, $k_{1\rho({23\cdots a} )}$ should be understood as 0 directly without using MCEL.}
\ba
{\rm Tadpoles ~in~}\eqref{twogeneral}=
&\label{tadpolegeneral}
\frac{1}{\ell^2 }
\sum_{a,b=1}^n
\phi_{\rho(a+1,\cdots, a) \mid  \sigma(b+1,\cdots, b) }
\prod_{p=1}^r (\ell-k_{1\rho({23\cdots a} )  })^{\mu_{p}}
\prod_{q=1}^t (\ell-k_{1\sigma({23\cdots b} )  })^{\nu_{q}}
  \,.
\ea
When there are no $\mu$ or $\nu$ indices, the above formula vanishes in a way similar to \eqref{u1cancel} but in general, it remains.
By identifying  $A_1=\rho(a+1,a+2,\cdots, a) $ and
${\tilde A}_1=\sigma(b+1,b+2,\cdots, b)$ and shifts of  loop momentum,  one can recognize the equivalence of the above equation and the tadpole part on the RHS of \eqref{twogeneral}.  For example, according to \eqref{twogeneral}, we have
\ba
\frac{1}{\ell^2}\int d\mu_2   {\rm PT}^\mu (1,2)   {\rm PT}^\nu(1,2)   \cong \,\,&
\ell^\mu\left(
\ell^\nu {\rm gon}(1,2)+ \frac{1}{s_{12}}\left[
\frac{\ell^\nu  }{\ell^2} - \frac{\ell^\nu-k^\nu}{\ell^2}
 - \frac{\ell^\nu+k^\nu}{\ell_1^2}  + \frac{\ell^\nu  }{\ell_1^2}
\right]
\right)\,,
\ea
while according to \eqref{tadpolegeneral}, it's given by
\ba
\label{eq519}
\ell^\mu\ell^\nu {\rm gon}(1,2)+\frac{1}{s_{12} \ell^2}\left[
(\ell^\mu-k_1^\mu)(\ell^\nu-k_1^\nu)-(\ell^\mu-k_1^\mu)\ell^\nu-\ell ^\mu(\ell^\nu-k_1^\nu)
+ \ell^\mu \ell^\nu
\right]\,.
\ea
Two representations are related by the shift of loop momentum $\ell_1\to \ell$.

As we have claimed before, whenever we mention shifts of loop momenta, we already used the MCEL to change the representation of loop propagators. Usually, it won't produce problems as the loop propagators themselves can never be singular. This is the reason why we include the tadpole contributions in the way shown in \eqref{twogeneral} where we use $\cong$. However, we emphasize that 
\eqref{tadpolegeneral} where we use $=$ is an exact identity. On the one hand, there are usually more singularities in tadpole terms. On the other hand, tadpoles are the only kind of topology whose quadratic propagators are obtained without using shifts of loop momenta, which makes the avoidance of using MCEL possible.  In the following, we explain the simplification of \eqref{tadpolegeneral} without using MCEL at all.

\subsubsection{Simplification}
The above formula \eqref{eq519} can be easily simplified and finally we get \eqref{v12v12} where the tadpole numerator is just $k_1^\mu k_1^\nu$ .  Inspired by this fact, we  continue to  simplify \eqref{tadpolegeneral} using the  Berends-Giele recursion of $\phi$ \cite{Mafra:2016ltu}. For rank-1 PT factors, i.e. $r=t=1$ in \eqref{tadpolegeneral},  we have a relatively simple result
\ba
\label{r=1t=1}
\frac{1}{\ell^2 s_{12\cdots n}}
\sum_{\substack{
(A_1,A_2)=(1,\rho)\\
(\tilde {A}_1, \tilde {A}_2)=(1,\sigma)
\\
\{A_1\}=\{{\tilde A}_1\}  \ni 1
}}
 \phi_{A_1|{\tilde A}_1}
  \phi_{ A_2|{\tilde A}_2}
\,
 \left( k_{A_1}^\mu~{\rm or}~- k_{A_2}^\mu
 \right)
  \left( k_{A_1}^\nu~{\rm or}~- k_{A_2}^\nu
 \right) \,,
\ea
where we choose $k_{A_1}^\mu$ if $\rho(n)\notin {A_1}$ or
$-k_{A_2}^\mu$ otherwise. Similarly, we choose $k_{A_1}^\nu$ if $\sigma(n)\notin {A_1}$ or
$-k_{A_2}^\nu$ otherwise.
We make these distinguishing in order to avoid the use of MCEL.   We see
in these cases, the numerators are always free of $\ell$.
We emphasize that even though we have used the 2-sequence  partitions here, we are still talking about tadpoles. The sign coming from $\phi \phi$ again becomes an overall sign ${\rm sgn}^\rho_\sigma$ shared with other polygons in \eqref{twogeneral}.   See \eqref{v123v132} as an example
and we complete the tadpoles in \eqref{v1234v1324}
 here,
\ba
\label{v1234v1324com}
&{\rm Tadpoles~in~}
\frac{1}{\ell^2}\int d\mu_4   {\rm PT}^{\mu} (1,2,3,4)   {\rm PT}^\nu(1,3,2,4)
\\
\non
=& \,\,
-
\frac{1}{\ell^2s_{1234}} \Bigg[
\frac{k_1^{\mu } k_1^{\nu }}{s_{23} s_{234}}+\frac{ k_{23}^{\mu } k_{23}^{\nu } }{s_{14}
   s_{23}}+\frac{k_{123}^{\mu } k_{123}^{\nu } }{s_{23}
   s_{123}}+\frac{k_3^{\mu } k_3^{\nu }}{s_{14}
   s_{124}}+\frac{k_2^{\mu } k_2^{\nu }}{s_{14} s_{134}}\Bigg]\,.
\ea

For general cases,  the tadpoles can be simplified as  %
\ba
&{\rm Tadpoles ~in~}\eqref{twogeneral}
\label{tadpolegeneralsim}
\\
=& \frac{1}{\ell^2 s_{12\cdots n}}
\sum_{\substack{
(A_1,A_2)=(1,\rho)\\
(\tilde {A}_1, \tilde {A}_2)=(1,\sigma)
\\
\{A_1\}=\{{\tilde A}_1\} \ni 1
}}
 \phi_{A_1| {\tilde A}_1}
  \phi_{A_2| {\tilde A}_2}
 \left(
 \prod_{p=1} ^r (\ell-k_{1{\bar A}A_2})^{\mu_p}
-
\prod_{p=1} ^r (\ell-k_{1{\bar A}})^{\mu_p}
\right)
\non
\\[-5mm]
&\qquad\qquad\qquad\qquad
\qquad\qquad\quad\,\,
\times
 \left(
 \prod_{q=1} ^t (\ell-k_{1 {\tilde {\bar A}}{ A}_2})^{\nu_q}
-
\prod_{q=1} ^t (\ell-k_{1{\tilde {\bar A}}})^{\nu_q}
\right)\,,
\non
  \ea
  where as usual, ${\bar A}$ and ${\tilde {\bar A}}$ are obtained by $A_1= A1{\bar A}$ and
${\tilde A}_1={\tilde  A}1{\tilde {\bar A}}$
respectively. The factor $\ell-k_{1{\bar A}A_2}$ comes from the coefficient of ${\rm PT}^{\rm tree}(+,A1{\bar A}A_2,- )$ in \eqref{ptvectorg}. So
  when $A=\{\}$, $k_{1{\bar A}A_2}$ should be understood as 0 directly. In this sense, we say the above formula is derived without using MCEL.  
  The details of the derivation
is put in appendix \ref{tadpolesim}. When $r=t=1$, the above formula reduces to \eqref{r=1t=1}.  The vanishing of tadpoles when $r$ or $t=0$ is also manifest.  Here we give some examples of \eqref{tadpolegeneralsim},
\ba
\frac{1}{\ell^2}\int d\mu_2   {\rm PT}^\mu (1,2)   {\rm PT}^{\nu\gamma}(1,2)  \cong &\,\,
\ell^\mu\ell^\nu \ell^\gamma {\rm gon}(1,2)+\frac{k_1^\mu (\ell^\nu k_1^\gamma +
k_1^\nu \ell^\gamma
- k_1^\nu k_1^\gamma   )}{\ell^2s_{12}}\,,
\\
\non
\frac{1}{\ell^2}\int d\mu_2   {\rm PT}^\mu (1,2)   {\rm PT}^{\nu\gamma\delta}(1,2)  \cong &\,\,
\ell^\mu\ell^\nu \ell^\gamma \ell^\delta {\rm gon}(1,2)+\frac{k_1^\mu \big(\ell^\nu   \ell ^\gamma  \ell^\delta-
 (\ell-k_1) ^\nu  (\ell-k_1) ^\gamma  (\ell-k_1) ^\delta 
 \big)
 }{\ell^2s_{12}}\,,
\ea
and complete the tadpoles in \eqref{vv123v132}
\ba
& \label{vv123v132com}
{\rm Tadpoles~in~}
\frac{1}{\ell^2}\int d\mu_3   {\rm PT}^{\mu\gamma} (1,2,3)   {\rm PT}^\nu(1,3,2)
\\
\non
=& \,\,
-
\frac{1}{\ell^2s_{123}} \Bigg[
\frac{(\ell^\mu k_1^\gamma +
k_1^\mu \ell^\gamma
- k_1^\mu k_1^\gamma   ) k_1^\nu}{s_{23}}
-\frac{
(\ell^\mu k_{12}^\gamma +
k_{12}^\mu \ell^\gamma
- k_{12}^\mu k_{12}^\gamma   )
 k_{3}^\nu}{s_{12}}
 \nl&
 \qquad\qquad
+\frac{ \big(
(\ell-k_{12})^\mu (\ell-k_{12})^\gamma
-(\ell-k_1)^\mu (\ell-k_1)^\gamma
 \big)
k_{13}^\nu}{s_{13}}
\Bigg]\,.
\non
\ea

\section{Towards the expansion onto basis \label{subcycle}}

Since we now understand what one-loop diagrams these PT factors produce, it is natural to ask how to expand one-loop CHY half-integrands onto them. At the tree level, plenty of papers have shown how to expand an arbitrary CHY half-integrand onto tree-level PT factors on the support of tree-level scattering equations \cite{Cardona:2016gon,Bjerrum-Bohr:2016axv,Du:2017gnh,Du:2017kpo,Huang:2017ydz,He:2018pol,He:2019drm,Edison:2020ehu}. The analog operations at the one-loop level are much more challenging as  the one-loop SE in \eqref{oneloopchy} usually just relate functions that cannot produce quadratic propagators.

It remains a {\it conjecture} that any quadratic half-integrand can be expanded onto generalized PT factors, and a systematic way to achieve this goal obviously deserves a lot of future efforts. In this note, we initiate these explorations by investigating a length-2 cycle
\ba
\label{bndef}
\frac{s_{12}\,{\rm PT}^{\rm tree}(12) }{
(-1)^{(n+1)}\s_3 \s_4\cdots \s_n} \cong
\frac{1}{(-1)^n \s_{12} \s_3\cdots \s_n } \left(
\frac{2\,\ell\! \cdot\! k_{2} }{\sigma_{2}}+\sum_{j=1 \atop j \neq 2}^{n} \frac{s_{2j} }{\sigma_{2 j}}
\right)
:= B_n(1,2)\,,
\ea
where ${\rm PT}^{\rm tree}(12)=-1/\s_{12}^2$.  We have used the SE in the first equality and defined this combination as $B_n(1,2)$. The first term in the above equation is very similar to the most symmetric factor \eqref{symmetricfactor} except that pair 1,2 is distinguished from the remaining points but 1 and 2 in the pair themselves are still symmetric.
Brute-force manipulations on the linear propagators produced by its CHY integral with a scalar PT factor at lower points  indeed lead to quadratic propagators, for example,
\ba
&\frac{1}{\ell^2}\int d\mu_4  B_4(1,2)  {\rm PT}(1,2,3,4)
\\
\cong&
\frac{\text{gon}(12,3,4) \left(2 \,\ell\cdot
   k_2-s_{23}-s_{24}\right)}{s_{12}}+\frac{1}{2} \text{gon}(1,2,3,4)
   \left(2\, \ell\cdot k_2+s_{12}-s_{23}-s_{24}\right)
   \nl
   &-\frac{2
   \left(s_{12}+s_{23}\right) \text{gon}(123,4)}{s_{12}
   s_{123}}+\frac{2 s_{24} \text{gon}(124,3)}{s_{12}
   s_{124}}-\text{gon}(1,23,4)
   \non\,.
\ea
Hence, as a first nontrivial check of the general conjecture, we should ask whether $B_n(1,2)$ can be expanded onto a generalized PT basis. A naive attempt to expand $B_n(1,2)$ without using any SE failed even though $B_n(1,2)$ itself is free of any cycle.  The next attempt, which uses the following ansatz with undetermined rational coefficients $x$'s and $y$'s
\ba
\sum_{\pi \in S_{n-1}}
\left[
\sum_{i=2}^n x_{i,\pi}
k_i^\mu {\rm PT}_\mu(1,\pi)
+
\sum_{2\leq i<j\leq n}^n y_{i,j,\pi}
s_{ij} {\rm PT}(1,\pi)
\right]\,,
\ea
still failed even though we have made use of SE and MCEL. On the assumption that there could be a pole in the coefficients of generalized PT factors, we finally found a way to expand it onto PT factors. For example (with $n\geq 4$),
\footnote{
In \eqref{bnexp}, there are no divergent propagators starting at $n=4$ so we can safely use MCEL. The generalization of \eqref{bnexp} to $n=3$ or even $n=2$ is possible if we apply  MCEL in a tricky way but we won't discuss them in this note.  }
\ba
\label{bnexp}
B_n(1,2)
\cong &
-\frac{1}{s_{12}}
\sum_{\pi\in S_{n-1}}
\Bigg[
4 \,k_{1,\mu}k_{2,\nu}L^{\mu\nu}_1+
\left(s_{12}
-\sum_{i=3}^n
{\rm sgn}^\pi _{2i}
\,
 s_{2,i} \right)
 \,
 k_{1,\mu} L_1^\mu
\Bigg]
{\rm PT}(1,\pi(2,3,\cdots,n))\,,
\ea
where we have used the operation defined in \eqref{operation}.   The above expansion is of course non-unique because the relevant tensorial, vectorial and scalar PT factors are not independent on the support of MCEL and SE. 

Based on some dualities between homology invariants and kinematic factors in superstring theory \cite{Mafra:2014gsa,Mafra:2017ioj}, one can also derive an identity that is essentially \eqref{bnexp} from the string side.
\footnote{We would like to thank Alex Edison for valuable discussions on these.} Here we comment that it is highly desirable to have a good understanding of this identity by using one-loop SE in \eqref{oneloopchy}. If that is available, it is highly plausible that we can find the generalization of \eqref{bnexp} to any cycle based on similar experience at the tree level. On the other hand, at the moment finding out all homology invariants at high multiplicities and high ranks is already very challenging from the string side, let alone dualities between them and kinematic factors. We leave the study of general expansions and possible connections with homology invariants from strings to the future.

\subsection{Expansion of $\ell_\mu \ell_\nu  \eta^{\mu \nu}$}

Starting from \eqref{bnexp}, we have already obtained various interesting expansions, such as that of $\ell_\mu \ell_\nu  \eta^{\mu \nu}$. Algebraically, we have
\ba
\frac{
\ell_\mu \ell_\nu  \eta^{\mu \nu}
}{(-1)^{n+1}\s_1\s_2\cdots \s_n}
=
\sum_{\pi\in S_{n-1}}
\Bigg[
\eta_{\mu\nu}
L^{\mu\nu}_1+{k_{1,\mu}L^\mu_1}
\Bigg]
{\rm PT}(1,\pi(2,3,\cdots,n))
+
\sum_{i=2}^n
B_n(1,i)
\,.
\ea
Together with \eqref{bnexp}, it gives rise to a nice expansion (with $n\geq 4$)
\ba
\label{expansionll}
&\frac{
\ell_\mu \ell_\nu  \eta^{\mu \nu}
}{(-1)^{n+1}\s_1\s_2\cdots \s_n}
\\ \non
\cong&
\sum_{\pi\in S_{n-1}}
\Bigg[
\left(
\eta_{\mu\nu}
-4\,\sum_{i=2}^n \frac{k_{1,\mu}k_{i,\nu}}{s_{1,i}}\right) L_1^{\mu\nu}
+
\!\!\!\!
\sum_{2\leq i<j\leq n}
\!\!\!\!
\left(\frac{1} {s_{1i}}-\frac{1} {s_{1j}}\right){\rm sgn}^\pi _{ij}
\,
s_{ij}
\,
k_{1,\mu} L_1^\mu
\Bigg]
{\rm PT}(1,\pi(2,3,\cdots,n))\,.
\ea
In the above two equations, we write $\ell_\mu \ell_\nu  \eta^{\mu \nu}$ in the numerators so that it's distinguished from the loop propagator  $\ell^2$ given in \eqref{loopintegral}.  In dimension regularization,   $\ell^2$ would be treated to live in $D-2\varepsilon$ dimensions while $\ell_\mu \ell_\nu  \eta^{\mu \nu}$ still lives in  $D$ dimensions. Hence their difference may produce some non-trivial results.
We remark that such expansion has immediate applications to {\it e.g.} the study of cancellation of the anomaly in heterotic strings (or even type I/II  superstrings) and their applications to field-theory amplitudes (see \cite{He:2017spx,Porkert:2022efy} for related discussions), since the only non-vanishing gauge variation of the parity odd part of their 6-pt correlators contains a term proportional to $\ell_\mu \ell_\nu  \eta^{\mu \nu}$. 
Let us end the section with an example of an $n=4$ CHY integral with $\ell_\mu \ell_\nu  \eta^{\mu \nu}$ which indeed gives quadratic propagators,
\ba
&\frac{1}{\ell^2}\int d\mu_4  \frac{
\ell_\mu \ell_\nu  \eta^{\mu \nu}
}{(-1)\s_1\s_2 \s_3 \s_4} \,{\rm PT}(1,2,3,4)
\cong {\rm gon}(1,2,3,4) \ell_\mu \ell_\nu  \eta^{\mu \nu}  - {\rm gon}^4(41,2,3)\,,
\ea
which contains an overall factor $\ell_\mu \ell_\nu  \eta^{\mu \nu}-\ell^2$ but integrates to a rational expression upon loop momentum in $D=6$ dimensions (see more in \cite{Porkert:2022efy}).

\section{Applications to physical amplitudes \label{applicationsec}}

One-loop CHY integrands in various theories have been organized as linear combinations of tree-level PT factors \eqref{treept}, for example for supersymmetric YM, GR, and EYM with various numbers of supercharges and their pure versions \cite{He:2017spx,Geyer:2017ela,Edison:2020uzf,Porkert:2022efy,Dong:2021qai}, which usually hold for arbitrary multiplicities and can be used to derive one-loop KLT-like relations and BCJ numerators with linear propagators.
Based on the general construction of forward limit in higher dimensions \cite{He:2015yua,Cachazo:2015aol}, more results for other theories  can be obtained similarly from their tree-level CHY formulae (see {\it c.f.} \cite{Cachazo:2014xea,He:2016iqi,Cachazo:2016njl,He:2018pol,Azevedo:2017lkz}).
Despite these successes, the absence of a systematic way to obtain quadratic propagators has blocked further progress for a long time. 
This has changed with the paper 
\cite{Edison:2021ebi} where homology invariants are used to produce quadratic propagators of 5-pt SYM and SG with maximal supersymmetry, which suggests generalization to higher points (though much more complicated).  In this section, we show the worldsheet correlators of SYM and SG can be expanded onto one-loop generalized PT factors \eqref{ptvectorg} at least for low multiplicities, which 
strongly suggests that higher-point correlators and those in other theories can be expanded similarly. A nice example would be 5-pt SYM and SG with maximal supersymmetry, where bubbles and triangles are absent. 
However, we choose to present less supersymmetric examples which are non-zero even for 3 points.  We take these examples as explicit applications of our general formulae to one-loop amplitudes in physical theories.

Our main object in this section is the 3-pt correlator with half-maximal supersymmetry, which has been used to construct loop integrands of half-maximally supersymmetric YM or GR.
\footnote{Here we refer to supergraivty constructed from double copy of $({\cal N}=2~{\rm SYM}) \times ({\tilde {\cal N}}=2~{\rm SYM}) $.}
A relatively compact form of the $n=3$ correlator is given in \cite{He:2017spx},
\begin{align}
\label{k3given}
\mathcal{K}_{3}^{1 / 2}=&\frac{1}{\s_1\s_2  \s_3}   \left( \ell_{\mu} C_{1 \mid 2,3}^{\mu, 1 / 2}+\frac14 s_{23} \, C_{1 \mid 23}^{1 / 2} \,  \frac{\sigma_2+\sigma_3}{\sigma_{2,3}} \right) \,,
\end{align}
where the exact definitions of $C_{1 \mid 2,3}^{\mu, 1 / 2}$ and $C_{1 \mid 23}^{1 / 2}$ can be found in that paper and we summarize them in the appendix \ref{appeappde}.   These two blocks are not independent and are related by
\ba
\label{relationsC}
-k_{2,\mu} C_{1|2,3}^{\mu,1/2} = k_{3,\mu} C_{1|2,3}^{\mu,1/2}= \frac12  s_{23} C_{1|23}^{1/2}\,.
\ea
On the support of these relations,
 the correlator can be algebraically expanded onto generalized PT factors,
\ba
\label{3pthalfexp}
\mathcal{K}_{3}^{1 / 2}
=
C_{1|2,3}^{\mu,1/2}  \Big( {\rm PT}_\mu(1,2,3)  + {\rm PT}_\mu(1,3,2) \Big)
-\frac{1}{4} s_{23} C_{1|23}^{1/2}   \Big( {\rm PT}(1,2,3)  - {\rm PT}(1,3,2) \Big)\,.
\ea
Once the above expansion is done, the quadratic-propagator representations of loop integrands for relevant theories are obtained straightforwardly as shown below. 

\subsection{Half-maximally supersymmetric YM}
Together with a scalar PT factor encoding the color ordering, the one-loop CHY integral of the correlator ${\cal K}^{1/2}_3$ gives the color ordered loop integrand of SYM:
\ba
{\cal M}^{{\rm SYM},1/2}(1,2,3)=&
\frac{1}{\ell^2}\int d\mu_3  {\rm PT} (1,2,3) \,
\mathcal{K}_{3}^{1 / 2}
\\
  \cong & \,
{\rm gon}(1,2,3)(C_{1|2,3}^{\mu,1/2} \ell_{\mu}-\frac{1}{4} s_{23} C_{1|23}^{1/2} )  + \frac{{\rm gon}(1,23) }{s_{23}}(-\frac{1}{2}  s_{23} C_{1|23}^{1/2} )  )
\nl&
+ \frac{{\rm gon}(12,3) }{s_{12}}\left((C_{1|2,3}^{\mu,1/2} \ell_{\mu}-\frac{1}{4} s_{23} C_{1|23}^{1/2} ) -(C_{1|2,3}^{\mu,1/2} [\ell_{\mu}+k_{2,\mu}]+\frac{1}{4} s_{23} C_{1|23}^{1/2} ) \right)
\nl&
+ \frac{{\rm gon}(31,2) }{s_{13}}
\left((C_{1|2,3}^{\mu,1/2} \big[ \ell_{\mu}+ k_{3,\mu} ] -\frac{1}{4} s_{23} C_{1|23}^{1/2} ) -(C_{1|2,3}^{\mu,1/2} \ell_{\mu} +\frac{1}{4} s_{23} C_{1|23}^{1/2} ) \right)
\,,
\non
\ea
where we have plugged in the results of  \eqref{s123s123}, \eqref{s1234s1243} and \eqref{s123123new}. On the support of \eqref{relationsC}, the above formula
simplifies to
\ba
\label{loopintegrandym}
{\cal M}^{{\rm SYM},1/2}(1,2,3)
  \cong & \,
{\rm gon}(1,2,3)(C_{1|2,3}^{\mu,1/2} \ell_{\mu}-\frac{1}{4} s_{23} C_{1|23}^{1/2} )
-\frac{1}{2} C_{1|23}^{1/2}\,  {\rm gon}(1,23)
\,,
\ea
which is consistent with the results in the literature \cite{Berg:2016fui}. 
We see that there are triangles and bubbles but no tadpoles because of the scalar PT factor. Besides, the 
scalar triangle term ${\rm gon}(1,2,3) s_{23} C_{1|23}^{1/2}$
can be dropped because of the vanishing Mandelstam variable. However, $s_{23} C_{1|23}^{1/2}$ indeed contributes to the bubble ${\rm gon}(1,23)$ by cancelling a divergent propagator $s_{23}^{-1}$.

\subsection{Half-maximally supersymmetric GR}
As for SG with half-maximal supersymmetry,
we need two copies of ${ {\cal K}}_3^{1/2}$ in the CHY integral.  Let's denote another copy as ${\hat {\cal K}}_3^{1/2}$ where  $C$ blocks of any rank are replaced by their hatted ones to indicate that they may contain different polarizations.
Then  the 3pt loop integrand for
half-maximally supersymmetric GR reads
\ba
\label{eq812}
{\cal M}^{{\rm SG},1/2}_3=\frac{1}{\ell^2}\int d\mu_3  \,
\mathcal{K}_{3}^{1 / 2} {\hat {\cal K}}_{3}^{1 / 2}
\,.
\ea
This time, the reason for the absence of tadpoles is no longer trivial.
The possible tadpole parts in the above equation are given by
\ba
{\rm Tadpoles~in~}
{\cal M}^{{\rm SG},1/2}_3
=
 {\rm Tadpoles~in~}
\frac{1}{\ell^2}\int d\mu_3
&C_{1|2,3}^{\mu,1/2}
\Big[
 {\rm PT}_\mu(1,2,3)
 +
 {\rm PT}_\mu(1,3,2) \Big]\,
 \\ \non&
 \times {\hat C}_{1|2,3}^{\nu,1/2}
 \Big[
 {\rm PT}_\nu(1,2,3)
 +
 {\rm PT}_\nu(1,3,2) \Big]\,.
 \ea
Plugging in the formulae
 \eqref{v123v132} and \eqref{v123v123}, we get
 \ba
&{\rm Tadpoles~in~}
{\cal M}^{{\rm SG},1/2}_3
\\
=&
 C_{1|2,3}^{\mu,1/2}   {\hat C}_{1|2,3}^{\nu,1/2}   \frac{1}{\ell^2s_{123}} \Bigg[
 \left(
\frac{k_{1,\mu} k_{1,\nu}}{s_{23}}
+\frac{k_{2,\mu} k_{2,\nu}}{s_{13}}
+\frac{k_{12,\mu} k_{12,\nu}}{s_{12}}
\right)+ \left(
\frac{k_{1,\mu} k_{1,\nu}}{s_{23}}
+\frac{k_{13,\mu} k_{13,\nu}}{s_{13}}
+\frac{k_{3,\mu} k_{3,\nu}}{s_{12}}
\right)
\nl
&- \left(
\frac{k_{1,\mu} k_{1,\nu}}{s_{23}}
-\frac{k_{2,\mu} k_{13,\nu}}{s_{13}}
-\frac{k_{12,\mu} k_{3,\nu}}{s_{12}}
\right)
-
\left(
\frac{k_{1,\mu} k_{1,\nu}}{s_{23}}
-\frac{k_{13,\mu} k_{2,\nu}}{s_{13}}
-\frac{k_{3,\mu} k_{12,\nu}}{s_{12}}
\right)
\Bigg]
\nl
=&C_{1|2,3}^{\mu,1/2}   {\hat C}_{1|2,3}^{\nu,1/2}   \frac{k_{123,\mu} \, k_{123,\nu}}{\ell^2s_{123}}  \left(
\frac{1}{s_{13}}
+\frac{1}{s_{12}}
\right)
\,.
 \ea
 By suitable applications of MCEL,  one can show that the above tadpole contribution vanishes.

The non-vanishing contribution in \eqref{eq812} is given by,
 \ba
 \label{eq79}
 {\cal M}^{{\rm SG},1/2}_3
  \cong & \,
{\rm gon}(1,2,3)(C_{1|2,3}^{\mu,1/2} \ell_{\mu}-\frac{1}{4} s_{23} C_{1|23}^{1/2} )  ({\hat C}_{1|2,3}^{\nu,1/2} \ell_{\nu}-\frac{1}{4} s_{23} {\hat C}_{1|23}^{1/2} )
\\
&+{\rm gon}(1,3,2)(C_{1|2,3}^{\mu,1/2} \ell_{\mu}+\frac{1}{4} s_{23} C_{1|23}^{1/2} )  ({\hat C}_{1|2,3}^{\nu,1/2} \ell_{\nu}+\frac{1}{4} s_{23} {\hat C}_{1|23}^{1/2} )
\nl&
+ \frac{1}{4}  {\rm gon}(1,23) \, s_{23}\,   C_{1|23}^{1/2}   {\hat C}_{1|23}^{1/2}
\,,
\non
\ea
once we have plugged in the result of
  \eqref{v123v123} in addition to  those of \eqref{s123s123}, \eqref{s1234s1243} and \eqref{s123123new}. Note that the two triangles on the RHS only differ by the positions of the loop momentum. By making use of
\ba
{\rm gon}(1,3,2) \big|_{\ell\to -\ell} ={\rm gon}(2,3,1),\qquad {\rm gon}(1,3,2) \big|_{\ell\to -\ell-k_1} ={\rm gon}(1,2,3)\,,
\ea
the loop integrand  \eqref{eq79} simplifies to
 \ba
 \label{loopintegrandgr}
 {\cal M}^{{\rm SG},1/2}_3
  \cong & \,
2 \, {\rm gon}(1,2,3)(C_{1|2,3}^{\mu,1/2} \ell_{\mu}-\frac{1}{4} s_{23} C_{1|23}^{1/2} )  ({\hat C}_{1|2,3}^{\nu,1/2} \ell_{\nu}-\frac{1}{4} s_{23} {\hat C}_{1|23}^{1/2} )
\\
&
+ \frac{1}{4}  {\rm gon}(1,23)
\,s_{23}\,  C_{1|23}^{1/2}   {\hat C}_{1|23}^{1/2}
\,.
\non
\ea
which is also consistent to the results in the literature \cite{Berg:2016fui} once we throw away the vanishing scalar terms  $\,s_{23}\,  C_{1|23}^{1/2} $ or $\,s_{23}\,  {\hat C}_{1|23}^{1/2} $. The loop integrals of triangles or bubbles have already been known for a long time in textbooks (see \cite{smirnov2013analytic} for example). Hence the loop integrals of \eqref{loopintegrandym} and \eqref{loopintegrandgr} and their UV divergence can be obtained straightforwardly. 


\paragraph{More comments:} Compared with the results in \cite{Edison:2021ebi}, one can see that 
 the 3pt cases with half-maximal supersymmetry are very similar to those of 5pt cases with maximal supersymmetry. This is a general pattern. 
The 4pt correlator with half-maximal supersymmetry is also very similar to that of
6pt correlator with maximal supersymmetry,  both of which are organized compactly in \cite{He:2017spx}. We claim that they can be algebraically expanded onto generalized PT factors and $B_n(1,i)$ defined in \eqref{bndef}. The latter can be expanded onto the generalized PT factor basis again on the support of SE in \eqref{oneloopchy} and MCEL according to \eqref{bnexp}. Hence the correlators of 4pt and 6pt can also be expanded onto a generalized PT basis. While quadratic-propagator representations have also been obtained using other methods \cite{Mafra:2014gja,Bridges:2021ebs,Berg:2016fui}, such results for 6-pt (4-pt) (half-)maximally supersymmetric GR can be produced from one-loop CHY formulae. We remark that the kinematic numerators needed in their cases are more involved and we will present them in future papers together with possible applications to higher points and more general theories.

\section{Conclusion and Discussions
\label{sec:con}}

In this paper, we proposed that one-loop generalized Parke-Taylor factors can serve as a basis of worldsheet functions that via one-loop CHY integrals produce one-loop diagrams with quadratic propagators.  The main result is our closed formula \eqref{twogeneral} for the CHY integral of any two generalized PT factors, where cyclic partitions are introduced to characterize the polygons and possible sub-trees. This is the one-loop generalization of the original formula for bi-adjoint $\phi^3$ amplitudes~\cite{Cachazo:2013iea}
. 
We also initiated the study of expanding worldsheet functions onto a generalized PT basis; as the first step towards applications for physical amplitudes, we have presented the result for three-point loop integrands in half-maximally supersymmetric YM theory and its double copy to supergravity. 

There are numerous open questions raised by our preliminary studies. First of all, it would be highly desirable to apply generalized PT factors to one-loop formulae for $n$-point amplitudes in various theories, and especially explore color-kinematic duality and double copy based on diagrams with quadratic propagators (see  \cite{Berg:2016fui,Mizera:2019blq,Bridges:2021ebs,He:2015wgf} for other related discussions using worldsheet methods); it also becomes easier to perform loop integration and study structure of integrated amplitudes. Relatedly, it would be very interesting to have a systematic way of expanding general quadratic half-integrands onto PT factors, which would allow us to prove our conjecture and have a better understanding of this basis. 

Certain half-integrands of physical theories, especially those without supersymmetry, might not be quadratic ones, but produce diagrams with quadratic propagators up to some terms homogeneous in $\ell$ which can be dropped since their loop integrals are scaleless \cite{He:2015yua,Cachazo:2015aol,Geyer:2017ela,Baadsgaard:2015twa}.  Our formulae should still be able to simplify the study of this possible kind of half-integrands by introducing another kind of equivalence rule in addition to SE and MCEL  to reduce them to PT basis.
Besides, it would be interesting to think about the possible modifications of our formulae if we consider the contributions of regular and singular solutions of the one-loop scattering equations separately, which may help us to regulate forward limits at the integrand level. It would also be interesting to study the geometric meaning of these worldsheet functions, especially in the context of scattering forms via scattering equations which have been studied extensively at tree level~\cite{Arkani-Hamed:2017mur,Herderschee:2019wtl,He:2018pue}.

Another class of worldsheet functions that produce quadratic propagators~\cite{Edison:2021ebi} are given by field-theory limits of homology invariants from string amplitudes. It would be very intriguing to establish relations of PT factors to those functions, which would provide a link to $\alpha'$ corrections, and string amplitudes via $Z$ and $J$ integrals~\cite{Zfunctions,Stieberger:2014hba}. Last but not least, the most exciting avenue with lots of data from (ambitwistor) string theory is higher loops (even with low multiplicities)~\cite{ Geyer:2016wjx,Geyer:2018xwu,Geyer:2019hnn,Geyer:2021oox, DHoker:2020prr,DHoker:2020tcq,DHoker:2021kks}, and we hope to understand more systematically how to derive loop diagrams and amplitudes with quadratic propagators from the worldsheet. 

\section*{Acknowledgement}
It is our pleasure to thank Alex Edison, Oliver Schlotterer, and Fei Teng for inspiring discussions and collaboration on related topics, and especially Oliver Schlotterer for valuable comments on the draft. YZ would like to thank Freddy Cachazo for his useful discussions. BF's research is supported by Chinese NSF funding under Grant No.11935013, No.11947301,  and No.12047502 (Peng Huanwu Center).  SH's research is supported in part by the Key Research Program of the Chinese Academy of Sciences, Grant NO. XDPB15, and by the National Natural Science Foundation of China under Grant No. 11935013,11947301, 12047502,12047503. The research of YZ was supported in part by a grant from the Gluskin Sheff/Onex Freeman Dyson Chair in Theoretical Physics and by Perimeter Institute. Research at Perimeter Institute is supported in part by the Government of Canada through the Department of Innovation, Science and Economic Development Canada and by the Province of Ontario through the Ministry of Colleges and Universities.

\appendix

\section{Proof of the main proposal (\ref{twogeneral}) \label{proofpro}}

Given that the generalized PT factor is  a sum of $n$ tree-level PT factors \eqref{ptvectorg}, the one-loop CHY integral in \eqref{twogeneral}
actually becomes $n^2$ terms of tree-level CHY integrals \eqref{oneloopchyori}.  Let's take a look at these tree-level integrals at first and then prove the general formula \eqref{twogeneral}.

\subsection{The intersection of two tree-level PT factors}

 To have a good picture for what we will get, let us focus a particular term like (identify $\rho(1)=1, \sigma(1)=1 $)
\ba {\rm PT}^{\rm tree} (+,\rho(i),\rho(i+1),...,\rho(i-1),-)\times
{\rm PT}^{\rm tree} (+,\sigma(j),\sigma(j+1),...,\sigma(j-1),-)\,.~~~\label{exp-3}
\ea
For this term, it is well known that the final result is the intersection of two color-ordered trivalent
Feynman diagrams with proper sign: one with the ordering $(+,\rho(i),\rho(i+1),...,\rho(i-1),-)$ and
another one  with the ordering $(+,\sigma(j),\sigma(j+1),...,\sigma(j-1),-)$. However,
for our purpose, we like to give a more clear organization of these color ordered trivalent
diagrams. Noticing that both tree-level PT factor has the same $(+,-)$ at the two ends, we would like to  classify these diagrams according to the structure of the line connecting the $+,-$ legs, i.e.,  how many internal  propagators  there are along the line:
\begin{itemize}

\item The first case is that there is no internal line, i.e.,  $+,-$ are connected by a cubic vertex, and all other legs are connected through the propagator with momentum $\sum_{i=1}^n k_i=k_{1...n}$. If we do not take the forward limit, it is not divergent. But if we take the forward limit, it will contribute to the tadpole. For this case,  diagrams can be summarized as
    \ba
     \phi_{\rho(i),\rho(i+1),...,\rho(i-1)|
    \sigma(j),\sigma(j+1),...,\sigma(j-1)
    }\,,~~~\label{exp-4-1}
    \ea
    The above expression has been used in \eqref{tadpolegeneral} but we present it here again as a starting point of general demonstration.
    It is important that with different choices of $i$ or $j$, the double current $\phi$ is different since the leg
    $-k_{1\cdots n }$ breaks the cyclic symmetry.
This is the key reason why we can use the $U(1)$ identity \eqref{u1cancel} to show the cancellation of tadpoles for certain situations.

    \item The second case is that there is one internal propagator. In this case, $n$ legs have been
    divided into two parts which in the ordering of the first PT factor read
    \ba A_1= [\rho(i),\rho(i+1),...,\rho(i+k) ],~~~~A_2=[ \rho(i+k+1),..,\rho(i-1)]\,,~~~\label{exp-4-2}
    \ea
  while  in the ordering of the second PT factor read
     \ba
    {\tilde A}_1= [\sigma(j),\sigma(j+1),...,\sigma(j+k) ],~~~~
    {\tilde A}_2=[ \rho(j+k+1),..,\rho(i-1)]\,.~~~\label{exp-4-2}
    \ea
    For this case to have nontrivial contribution, we must require $\{A_i\}=\{{\tilde A}_i\}$, $i=1,2$
    and the result  can be summarized as
    \ba {1\over s_{A_1,\ell}}
    & \phi_{\rho(i),\rho(i+1),...,\rho(i+k)|
    \sigma(j),\sigma(j+1),...,\sigma(j+k)
    }
    \,
     \phi_{\rho(i+k+1),..,\rho(i-1)|
     \sigma(j+k+1),..,\sigma(i-1) }\,.~~~\label{exp-4-3}\ea

    \item Now the pattern is clear. For the line having $m-1$ internal propagators, we should
    divide the ordering $(\rho(i),\rho(i+1),...,\rho(i-1))$ to $m$ sequences $A_1,A_2,\cdots, A_m$.
    So does the second ordering which gives ${\tilde A}_1,{\tilde A}_2,\cdots, {\tilde A}_m$ where $A_j$ and ${\tilde A}_j$ are the same up to a permutation of their elements for all $j=1,2,\cdots,m$.
    For each pair of
    sequences $A_i, {\tilde A}_i$, the contribution to the corner is $\phi_{A_i\mid\tilde{A}_i}$. When the sequence has only one element,
    the contribution is just $1$.
    Plugging back the propagators along the line we have
    \ba \left (\prod_{i=1}^{m-1} {1\over s_{A_{12\cdots i},\ell}}\right)
     \left (\prod_{i=1}^{m}
    \phi_{A_i\mid\tilde{A}_i}
    \right)\,.
    ~~~\label{exp-4-4}
    \ea
    A particular case is when all sequences have only one element, which will contribute to the $n$-gon.

\end{itemize}

Having the picture of tree-level integrals, let's continue to prove \eqref{twogeneral}.

\subsection{Proof}
Without loss of generality,  let's take rank-1 PT factors as examples, i.e, $r=t=1$ in \eqref{twogeneral}.

  \begin{proof}
    Since the tadpole part have been explained in \eqref{tadpolegeneral}, we just need to consider the cases with $2\leq m\leq n $ in the formula \eqref{twogeneral}.
Let's see what kind of terms could contribute to a given $m$-gon when we expand the one-loop CHY integrals on the LHS of  \eqref{twogeneral}
as sums of tree-level ones.   Obviously,
 only the following $m$  tree-level integrals
\ba
\label{twogeidpr1}
\frac{1}{\ell^2}\sum_{i=1}^m \int d\mu_{n+2}^{\rm tree}&
{\rm PT}^{\rm tree}(+,A_{i+1},A_{i+2},\cdots, A_{i},-)  (\ell^\mu-k_{1{\bar A},A_{2\cdots i}}^\mu)
\nl&
\times
 {\rm PT}^{\rm tree}(+,{\tilde A}_{i+1},{\tilde A}_{i+2},\cdots, {\tilde A}_{i},-)(\ell^\nu-k_{1{\bar A},A_{2\cdots i}}^\nu)  \,,
\ea
could contribute to the $m$-gon ${\rm gon}^A(A_1, \cdots, A_m)$ on the RHS of \eqref{twogeneral}.
What's more,
the relevant tree-level trivalent graphs they produce are given by
\ba
\label{twogeidpr3}
\left(
\sum_{i=1}^m
\frac{
(\ell^\mu-k_{1{\bar A},A_{2\cdots i}}^\mu)
(\ell^\nu-k_{1 \tilde{{\bar A}},A_{2\cdots i}}^\nu)
 }{\ell^2 s_{A_{i+1},\ell } s_{A_{i+1,i+2},\ell }
\cdots
s_{A_{i+1,i+2,\cdots, i-1},\ell }
}
\right)
\prod_{j=1}^m \phi_{A_j\mid\tilde{A}_j}\,,
\ea
while all other trivalent graphs with linear loop propagators they produce would contribute to other polygons.   Rewriting
\ba
\ell^\mu-k_{1{\bar A},A_{2\cdots i}}^\mu =\ell^\mu-k_{A_{12\cdots i}}^\mu + k_{A}^\mu,
\qquad
\ell^\nu- k_{1 \tilde{{\bar A}},A_{2\cdots i}}^\nu  =\ell^\mu- k_{A_{12\cdots i}}^\nu  + k_{{\tilde A}}^\nu\,,
\ea
and then 
using partial fraction and shifts of loop momenta like \eqref{partialgeneral2}, one can recognize the summation in the big parenthesis of the above equation assembles to the $m-$gon,
\ba
(\ell^\mu+k_{A}^\mu)(\ell^\nu+k_{\tilde A}^\nu) {\rm gon}(A_1,\cdots ,A_m) \cong \ell^\mu (\ell^\nu-\Delta^\nu_{A{\tilde A}}) {\rm gon}^A\, (A_1,\cdots ,A_m)\,,
\ea
The $\ell$-dependent numerator, $\ell^\mu$, on the RHS of the above equation becomes an overall factor for all polygons and
can be extracted out.
Together with the remaining factors in \eqref{twogeidpr3}, we produced the exact $m$-gon terms on the RHS of \eqref{twogeneral},
\ba
\ell^\mu (\ell^\nu-\Delta^\nu_{A{\tilde A}}) {\rm gon}^A\, (A_1,\cdots ,A_m) \prod_{j=1}^m \phi_{A_j\mid\tilde{A}_j}\,.
\ea
Up to now, we can say
any tree-level trivalent graph that contains any linear loop propagators produced on the LHS of \eqref{twogeneral} would contribute to a particular polygon on the RHS of \eqref{twogeneral}.  Besides, they succeed in making up all kinds of polygons with proper numerators.  Hence together with the tadpole contributions explained in section \ref{sectiontadpole},  we proved \eqref{twogeneral}. Since \eqref{scalarid} and \eqref{scalardi} are special cases of it, we proved them as well.

\end{proof}

\section{\label{tadpolesim}Derivation  of (\ref{tadpolegeneralsim})
}

We first recall the Berends-Giele recursion
of bi-adjoint $\phi^3$ theory  double-current  \cite{Mafra:2016ltu}
\begin{equation}
    \phi_{\rho(a+1, \cdots, a) \mid\sigma(b+1, \cdots, b)}
   =\frac{1}{s_{1,2,\dots,n}}\sum_{\substack{
XY=\rho(a+1,\cdots,a)
\\
PQ=\sigma(b+1,\cdots, b)
}}
\left(\phi_{X \mid P} \phi_{ Y \mid Q}-
\phi_{X \mid Q} \phi_{ Y \mid P}
\right),
\end{equation}
where $\{X, Y\}$ is  a partition of the sequence $\rho(a+1,\cdots,a)$, which cannot be empty.  $\phi_{X \mid P}$ vanishes if $X$ and $P$ are not the same up to a permutation of their elements. In particular, we have  $\phi_{i|j}=\delta_{i,j}$.
The above recursion allows us to rewrite \eqref{tadpolegeneral} as
\ba
& {\rm Eq.} \eqref{tadpolegeneral}
\label{tadpoleBG}
\\
=&\frac{1}{\ell^2 s_{1,2,\dots,n}}
\sum_{a,b=1}^n
\sum_{\substack{
XY=\rho(a+1,\cdots,a) \\
PQ=\sigma(b+1,\cdots, b)
}}
\!\!\!
\left(\phi_{X \mid P} \phi_{Y \mid Q}-\phi_{X \mid Q} \phi_{ Y \mid P} \right) \prod_{p=1}^r (\ell-k_{1\rho({23\cdots a} )  })^{\mu_{p}}
\prod_{q=1}^t (\ell-k_{1\sigma({23\cdots b} )  })^{\nu_{q}}
 \nl
 =
&\frac{1}{\ell^2 s_{1,2,\dots,n}}
\sum_{\substack{
(X,Y)=(1,\rho(2,\cdots,n)) \\
(P,Q)=(1,\sigma(2,\cdots, n))
}}
\!\!\!
\left(\phi_{X \mid P} \phi_{Y \mid Q}-\phi_{X \mid Q} \phi_{ Y \mid P} \right) \prod_{p=1}^r (\ell-k_{ [XY]_1  })^{\mu_{p}}
\prod_{q=1}^t (\ell-k_{[PQ]_1  })^{\nu_{q}}
 \,,
 \non
\ea
where in the last line we have rearranged  summations  and  treat $\{X,Y\}$ as a {\it cyclic} partition of $(\rho(a+1,\cdots,a))=(1,\rho(2,\cdots, n))$.
 Note that both $X$ or $Y$ can contain ${\rho(1)}=1$.  We define $[...]_1$ acting on a sequence as its subsequence starting from 1 to the end.  For example,  $[\rho(a+1,\cdots,n), 1, \rho(2,\cdots,a)]_1 = 1 \rho(2,\cdots,a)$.   Using this,
 $k_{1\rho(23\cdots a)}$ in the second line can be written as  $k_{[XY]_1}$ in the third line once $\{X,Y\}$ is chosen.

Now let's see what kind of $X,Y,P,Q$ can produce
 a certain term
 $\phi_{A_1|{\tilde A}_1}\phi_{A_2|{\tilde A}_2}$ for given
 cyclic partitions $(A_1, A_2)=(1,\rho(2,\cdots, n)),(\tilde A_1, \tilde A_2)=(1,\sigma(2,\cdots, n)) $ with $A_1, \tilde A_1\ni 1$. It's easy to see that there are the following four choices,
\ba
\{X=A_1,Y=A_2,P=\tilde{A}_1,Q=\tilde{A}_2\},\quad
\{X=A_1,Y=A_2,P=\tilde{A}_2,Q=\tilde{A}_1\},
\nl
\{X=A_2,Y=A_1,P=\tilde{A}_1,Q=\tilde{A}_2\},\quad
\{X=A_2,Y=A_1,P=\tilde{A}_2,Q=\tilde{A}_1\}.
\ea

For $X=A_1,Y=A_2$, we have $[XY]_1=1{\bar A}A_2$, so there would be a factor
  $\prod_{p=1}^r\left(\ell-k_{1\bar{A}A_2}\right)^{\mu_p}$. As for $X=A_2,Y=A_1$, we have $[XY]_1=1{\bar A}$, so there would be a factor
  $\prod_{p=1}^r\left(\ell-k_{1\bar{A}}\right)^{\mu_p}$. Similar for $\{P=\tilde{A}_1,Q=\tilde{A}_2\}$ and $\{P=\tilde{A}_2,Q=\tilde{A}_1\}$.  Hence the four  choices together give rise to
 \ba
 \frac{1}{\ell^2 s_{12\cdots n}}
 \phi_{A_1| {\tilde A}_1}
  \phi_{A_2| {\tilde A}_2}
& \left(
 \prod_{p=1} ^r (\ell-k_{1{\bar A}A_2})^{\mu_p}
-
\prod_{p=1} ^r (\ell-k_{1{\bar A}})^{\mu_p}
\right)
 \left(
 \prod_{q=1} ^t (\ell-k_{1 {\tilde {\bar A}}{ A}_2})^{\nu_q}
-
\prod_{q=1} ^t (\ell-k_{1{\tilde {\bar A}}})^{\nu_q}
\right)\,.
  \ea
  Summing over all cyclic partitions of $(A_1, A_2)=(1,\rho(2,\cdots, n)),(\tilde A_1, \tilde A_2)=(1,\sigma(2,\cdots, n)) $ with $A_1, \tilde A_1\ni 1$ leads to \eqref{tadpolegeneralsim}.

\section{Integrals with two arbitrary orderings
\label{twomostgen}
}

Mention that
 the proposal \eqref{twogeneral} can also be written more symmetrically concerning both $\mu$ and $\nu$ indices by shifts of loop momenta at the cost of absences of overall $\ell^\mu$ factors and universal position of $\ell$ in every one-loop trivalent graph,
\ba
\label{twogeneralsym}
&\frac{1}{\ell^2}\int d\mu_n
{\rm PT}^{\mu_1,\cdots,\mu_r}
(1,\rho(2,\cdots,n))
 {\rm PT}^{\nu_1,\cdots,\nu_t}
(1,\sigma(2,\cdots,n))
 \\
 \non
\cong&
\!\!\!\!\!\!\!\!\!\!\!\!
\sum_{
\substack
{
(A_1,A_2,\cdots,A_m)=(1,{\rho}(2,\cdots,n)), \, A_1=A1{\bar A} \\
(\tilde A_1, \tilde A_2,\cdots, \tilde A_m)=(1,{\sigma}(2,\cdots,n)),\,{\tilde A}_1={\tilde A}1{\tilde {\bar A}}
\\
1\leq m\leq n, \quad \{A_j\}= \{\tilde A_j\} ~\forall 1\leq j\leq m
}}
\!\!\!\!\!\!\!\!\!\!\!\!
 \ell^{\mu_1}_A \cdots
\ell^{\mu_r} _A
 \ell^{\nu_1} _{\tilde A} \cdots
\ell^{\nu_t} _{\tilde A}\,
{\rm gon}(A_1, \cdots, A_m) \prod_{i=1}^m  \phi_{A_i\mid\tilde{A}_i}\,,
\ea
where $\ell_A^\mu=\ell^\mu+k_{A}^\mu$.  We have listed all of the hidden details in the summation of \eqref{twogeneral} here to avoid possible confusion as we are going to talk about the most general orderings. In this representation, there is an overall factor $1/\ell^2$.

For two arbitrary orderings ${\bm \rho}(1,2,\cdots,n)$ and ${\bm \sigma}(1,2,\cdots,n)$
where ${\bm \rho}(1)$ and  ${\bm \sigma}(1)$ may not equal to 1, we propose that
\ba
\label{twomostgeneral}
&\frac{1}{\ell^2}\int d\mu_n
{\rm PT}^{\mu_1,\cdots,\mu_r}
({\bm \rho}(1,2,\cdots,n))
 {\rm PT}^{\nu_1,\cdots,\nu_t}
({\bm \sigma}(1,2,\cdots,n))
 \\
 \non
\cong&
\!\!\!\!\!\!\!\!\!\!\!\!
\sum_{
\substack
{
(A_1,A_2,\cdots,A_m)=({\bm \rho}), \, A_1=A1{\bar A} \\
(\tilde A_1, \tilde A_2,\cdots, \tilde A_m)=({\bm \sigma}),\,{\tilde A}_1={\tilde A}1{\tilde {\bar A}}
\\
1\leq m\leq n, \quad \{A_j\}= \{\tilde A_j\} ~\forall 1\leq j\leq m
}}
\!\!\!\!\!\!\!\!\!\!\!\!
 \,
{\rm gon}(A_1, \cdots, A_m)
\left(
 \prod_{i=1}^m
 \phi_{A_i\mid\tilde{A}_i}
 \right) \left(  \prod_{p=1}^r
\left(
 \ell_A +k_{1,\cdots, {\bm \rho}({n-1}), {\bm \rho}({n})} \right) ^{\mu_p}
 \right)
 \\[-8mm]
&
\qquad \qquad  \qquad\qquad \qquad  \qquad\qquad \qquad  \qquad\qquad
\times
 \left(
 \prod_{q=1}^t
\left(
 \ell_{\tilde A} +k_{1,\cdots, {\bm \sigma}({n-1}), {\bm \sigma}({n})}  \right)  ^{\nu_q}
 \right)
\,,
\non
\ea
where the summation is almost the same as that in \eqref{twogeneral} or \eqref{twogeneralsym}
except that $(1,\rho(2,\cdots,n))$ is replaced by $({\bm \rho}(1,2,\cdots,n))$ in the cyclic partitions. Compared to \eqref{twogeneralsym}, we have additional shift factor,  $k_{1,\cdots, {\bm \rho}({n-1}), {\bm \rho}({n})}$, which means the sum of momenta between 1 (included) and ${\bm \rho}(1)$ (not included) in the cyclic ordering $( {\bm \rho}(1,2,\cdots, n))$. Supposing ${\bm \rho}(u)=1$, we have
\ba
k_{1,\cdots, {\bm \rho}({n-1}), {\bm \rho}({n})}= k_{{\bm \rho}(u)}+k_{{\bm \rho}( u+1)}+\cdots+ k_{{\bm \rho}( n-1)}+k_{{\bm \rho}( n)}\,.
\ea
When ${\bm \rho}(1)=1$,  $k_{1,\cdots, {\bm \rho}({n-1}), {\bm \rho}({n})}$ should be understood as 0 directly without using MCEL.
When ${\bm \rho}(1)={\bm \sigma}(1)=1$,  \eqref{twomostgeneral} reduces to \eqref{twogeneralsym}.

The tadpole contributions are also encoded in
\eqref{twomostgeneral}. Just like  \eqref{tadpolegeneral}, we can also get the tadpoles by just expanding the one-loop CHY integrals as tree-level ones and collecting all trivalent graphs where the legs $+$ and $-$ share a vertex,
\ba
{\rm Tadpoles ~in~}\eqref{twomostgeneral}=
&\label{tadpolegeneralmost}
\frac{1}{\ell^2 }
\sum_{a,b=1}^n
 \phi_{{\bm \rho}(a+1,\cdots, a) |{\bm \sigma}(b+1,\cdots, b)}
\prod_{p=1}^r (\ell-k_{{\bm \rho}({123\cdots a} )  })^{\mu_{p}}
\prod_{q=1}^t (\ell-k_{{\bm \sigma}({123\cdots b} )  })^{\nu_{q}}
 \,.
\ea

Here are some examples of \eqref{twomostgeneral},
\ba
\frac{1}{\ell^2}\int d\mu_2   {\rm PT}^\mu (2,1)   {\rm PT}(1,2)  \cong&\,
(\ell^\mu+k_1^\mu) {\rm gon}(1,2)\,,
\nl
\frac{1}{\ell^2}\int d\mu_3   {\rm PT}^\mu (3,1,2)   {\rm PT}(1,2,3)   \cong& \,\,
(\ell^\mu+k_{12}^\mu) \left[{\rm gon}(1,2,3)+ \frac{{\rm gon}(12,3) }{s_{12}}+ \frac{{\rm gon}(1,23) }{s_{23}}+ \frac{{\rm gon}^3(31,2) }{s_{13}}\right],
\nl
\frac{1}{\ell^2}\int d\mu_3   {\rm PT}^{\mu\nu}(2,3,1)  {\rm PT}(1,3,2)   \cong &
-
(\ell^\mu +k_1^\mu) (\ell^\nu+k_1^\nu)\left[\frac{{\rm gon}(12,3) }{s_{12}}+ \frac{{\rm gon}(1,23) }{s_{23}}+ \frac{{\rm gon}^3(31,2) }{s_{13}}
\right],
\nl
\frac{1}{\ell^2}\int d\mu_2   {\rm PT}^\mu (2,1)   {\rm PT}^\nu (1,2)  \cong&\,
(\ell^\mu+k_1^\mu) \ell^\nu {\rm gon}(1,2) -\frac{k_2^\mu k_1^\nu }{s_{12}}\,.
\label{v21v12}
\ea

The proof of \eqref{twomostgeneral}  is very similar to that of \eqref{twogeneral}. The main difference is that for a given $m$-gon, the possible contributions from tree-level CHY integrals are given by
\ba
\label{twogeidpr1}
\frac{1}{\ell^2}\sum_{i=1}^m \int d\mu_{n+2}^{\rm tree}&
{\rm PT}^{\rm tree}(+,A_{i+1},A_{i+2},\cdots, A_{i},-)
\prod_{p=1}^r (\ell-k_{1{\bar A},A_{2\cdots i}}+k_{1,\cdots, {\bm \rho}({n-1}), {\bm \rho}({n})} )^{\mu_p}
\nl&
\times
 {\rm PT}^{\rm tree}(+,{\tilde A}_{i+1},{\tilde A}_{i+2},\cdots, {\tilde A}_{i},-)
 \prod_{q=1}^t
 (\ell-k_{1{\bar A},A_{2\cdots i}}+k_{1,\cdots, {\bm \rho}({n-1}), {\bm \rho}({n})})^{\nu_t}  \,,
\ea
where we have seen the additional shift factors
$k_{1,\cdots, {\bm \rho}({n-1}), {\bm \rho}({n})}$ and $k_{1,\cdots, {\bm \sigma}({n-1}), {\bm \sigma}({n})}$, which are inherited by the final result \eqref{twomostgeneral}.

\section{\label{appeappde}Kinematic factors used in ${\cal K}_3^{1/2}$}
As shown in \cite{He:2017spx,Berg:2016fui},  the kinematic factors used in \eqref{k3given} are given by
\ba
C_{1 \mid 2,3}^{\mu, 1 / 2} :=  t_{1,2,3}^{\mu}+k_{2}^{\mu} t_{12,3}+k_{3}^{\mu} t_{13,2}\,,
\qquad
C_{1 \mid 23}^{1 / 2} :=  t_{1,23}+t_{12,3}-t_{13,2}\,,
\ea
with
\ba
t_{1,2}&=(\epsilon_1 \cdot \epsilon_2) (k_1 \cdot k_2) -(\epsilon_1 \cdot k_2) (\epsilon_2 \cdot k_1) ,\qquad
t_{12,3}=t_{3,12}=\left(\epsilon_{1} \cdot \epsilon_{2}\right)\left(k_{1} \cdot \epsilon_{3}\right),
\nl
t_{1,2,3}^{\mu}&=\epsilon_1^\mu  t_{2,3} +\epsilon_2^\mu  t_{3,1}+\epsilon_3^\mu  t_{1,2}\,,
\ea
where $\epsilon_i$ are polarizations.

\bibliographystyle{JHEP}
\bibliography{cites}{}

\end{document}